\documentclass[10pt,rmp,aps,superscriptaddress,notitlepage,twocolumn,showkeys]{revtex4-1}
\usepackage{graphicx}
\usepackage{amssymb}
\usepackage{amsmath}
\usepackage{amsthm}
\usepackage{mathrsfs}
\usepackage{url}
\usepackage{color}
\usepackage{longtable}

\definecolor{red}{rgb}{1,0,0}

\newtheorem*{mydef}{Definition}

\newcommand{\appsection}[1]{\let\oldthesection\thesection
  \renewcommand{\thesection}{Appendix \oldthesection}
  \section{#1}\let\thesection\oldthesection}

\begin{document}

\title{Limit order books}

\author{Martin D. Gould}
\email[Corresponding author: ]{gouldm@maths.ox.ac.uk}
\affiliation{Oxford Centre for Industrial and Applied Mathematics, Mathematical Institute, University of Oxford, Oxford OX1 3LB, UK}
\affiliation{Oxford-Man Institute of Quantitative Finance, University of Oxford, Oxford, OX2 6ED, UK}
\affiliation{CABDyN Complexity Centre, University of Oxford, Oxford OX1 1HP, UK}
\author{Mason A. Porter}
\affiliation{Oxford Centre for Industrial and Applied Mathematics, Mathematical Institute, University of Oxford, Oxford OX1 3LB, UK}
\affiliation{CABDyN Complexity Centre, University of Oxford, Oxford OX1 1HP, UK}
\author{Stacy Williams}
\affiliation{FX Quantitative Research, HSBC Bank, London E14 5HQ, UK}
\author{Mark McDonald}
\affiliation{FX Quantitative Research, HSBC Bank, London E14 5HQ, UK}
\author{Daniel J. Fenn}
\affiliation{FX Quantitative Research, HSBC Bank, London E14 5HQ, UK}
\author{Sam D. Howison}
\affiliation{Oxford Centre for Industrial and Applied Mathematics, Mathematical Institute, University of Oxford, Oxford OX1 3LB, UK}
\affiliation{Oxford-Man Institute of Quantitative Finance, University of Oxford, Oxford, OX2 6ED, UK}


\begin{abstract}Limit order books (LOBs) match buyers and sellers in more than half of the world's financial markets.  This survey highlights the insights that have emerged from the wealth of empirical and theoretical studies of LOBs.  We examine the findings reported by statistical analyses of historical LOB data and discuss how several LOB models provide insight into certain aspects of the mechanism.  We also illustrate that many such models poorly resemble real LOBs and that several well-established empirical facts have yet to be reproduced satisfactorily.  Finally, we identify several key unresolved questions about LOBs.\end{abstract}

\maketitle

\section{Introduction}

More than half of the markets in today's highly competitive and relentlessly fast-paced financial world now use a \emph{limit order book (LOB)} mechanism to facilitate trade \citep{Rosu:2009}.  The Helsinki, Hong Kong, Shenzhen, Swiss, Tokyo, Toronto, and Vancouver Stock Exchanges, together with Euronext and the Australian Securities Exchange, all now operate as pure LOBs \citep{Luckock:2001, Gu:2008empiricalregularities}; the New York Stock Exchange (NYSE), NASDAQ, and the London Stock Exchange (LSE) \citep{Cont:2010} all operate a bespoke hybrid LOB system.  Thanks to technological advances, traders worldwide have real-time access to the current LOB, providing buyers and sellers alike ``the ultimate microscopic level of description'' \citep{Bouchaud:2002}.

In an LOB, complicated global phenomena emerge as a result of the local interactions between many heterogeneous agents when the system throughput becomes sufficiently large.  This makes an LOB an example of a \emph{complex system} \citep{Mitchell:2009complexity}.  The unusually rich, detailed, and high-quality historic data from LOBs provides a suitable testing ground for theories about well-established statistical regularities common to a wide range of markets \citep{Bouchaud:2009, Cont:2001, Farmer:2003}, as well as for popular ideas in the complex systems literature such as universality, scaling, and emergence.

The many practical advantages to understanding LOB dynamics include: gaining clearer insight into how best to act in given market situations \citep{Harris:1996}; optimal order execution strategies \citep{Obizhaeva:2013}; market impact minimization \citep{Eisler:2012}; designing better electronic trading algorithms \citep{Engle:2006measuring}; and assessing market stability \citep{Kirilenko:2011flash}.  In this survey, we discuss some of the key ideas that have emerged from the analysis and modelling of LOBs in recent years, and we highlight the strengths and limitations of existing LOB models.

Investigations of LOBs have taken a variety of starting points, drawing on ideas from economics, physics, mathematics, statistics, and psychology.  Unsurprisingly, there is no clear consensus on the best approach.  This point is exemplified by the contrast between the approach normally taken in the economics literature, in which models focus on the behaviour of individual traders and present LOBs as sequential games \citep{Foucault:1999, Parlour:1998, Rosu:2009}, with the approach normally taken in the physics literature, in which order flows are treated as random and techniques from statistical mechanics are used to explore the resulting dynamics \citep{Challet:2001, Cont:2010, Smith:2003}.  In the present paper, we discuss developments in both the economics and physics literatures, and we emphasize aspects of LOBs that are most relevant to practitioners.

Several other survey articles focus on particular aspects of LOBs.  \citet{Friedman:1993} reviewed early studies of \emph{double auction} style trading, of which LOBs are an example.  \citet{Parlour:2008} addressed the economic and theoretical aspects of LOB trading.  \citet{Bouchaud:2009} assessed the current understanding of price formation in LOBs.  \citet{Chakraborti:2011a, Chakraborti:2011b} examined the role of econophysics in understanding LOB behaviour.  In the present survey we note the similarities and differences between several empirical studies of historical LOB data, discuss LOB models from both the physics and economics literatures, highlight several modelling assumptions that are not well-supported by the empirical findings, and identify several key unresolved questions.

The remainder of the survey is organized as follows.  In Section \ref{lobfunction}, we give formal definitions related to LOBs and formulate a mathematically precise description of LOB trading.  In Section \ref{LOMs}, we discuss some practical aspects of trading via LOBs and examine the difficulties that arise in quantifying them.  In Section \ref{empiricals}, we examine the important role of empirical studies of LOBs, highlighting both consensus and disagreement within the literature.  We examine a selection of models in Section \ref{themodel}.  In Section \ref{openproblems}, we discuss key unresolved problems, and we conclude in Section \ref{concl}.

\section{A Mathematical Description of an LOB}\label{lobfunction}

In this section, we formulate a precise description of trading that is common to most LOB markets.  Of course, some individual exchanges and trading platforms operate slight variations of these core principles.  \citet{Harris:2003trading} provides a comprehensive review of specific details governing particular exchanges, so we do not reproduce them here.

\subsection{Preliminaries}\label{prelims}

Before LOBs grew in popularity, most financial trades took place in \emph{quote-driven} marketplaces, in which a handful of large \emph{market makers} centralize buy and sell orders by publishing the prices at which they are willing to buy and sell the traded asset.  The market makers set their sell price higher than their buy price in order to earn a profit in exchange for providing \emph{liquidity}\footnote{Liquidity is difficult to define formally.  \citet{Kyle:1985} identified the three key properties of a liquid market to be tightness (``the cost of turning around a position over a short period of time''), depth (``the size of an order-flow innovation required to change prices a given amount''), and resiliency (``the speed with which prices recover from a random, uninformative shock'').} to the market, for taking on the risk of acquiring an undesirable inventory position, and for being exposed to possible \emph{adverse selection} (i.e., encountering other traders who have better information about the value of the asset and who can therefore make a profit by buying or selling, often repeatedly, with the market maker \citep{Parlour:2008}).  The only prices available to other traders who want to buy or sell the asset are those made public by the market makers, and the only action available to such traders is to immediately buy or sell at one of the market makers' prices.  Ticket touts exemplify a quote-driven market in action.

An LOB is much more flexible because every trader has the option of posting buy (respectively, sell) \emph{orders.}

\begin{mydef}An \emph{order $x=(p_x,\omega_x, t_x)$} submitted at time $t_x$ with price $p_x$ and size $\omega_x>0$ (respectively, $\omega_x<0$) is a commitment to sell (respectively, buy) up to $\left|\omega_x\right|$ units of the traded asset at a price no less than (respectively, no greater than) $p_x$.\end{mydef}

We introduce the vector notation $x=(p_x,\omega_x, t_x)$ because it allows explicit calculation of the \emph{priority} (see Section \ref{priority}) of any order at any time.

For a given LOB, the units of order size and price are set as follows.

\begin{mydef}The \emph{lot size} $\sigma$ of an LOB is the smallest amount of the asset that can be traded within it.  All orders\footnote{In some markets, there are two lot-size parameters: a minimum size $\sigma$ and an increment $\varepsilon$.  In such markets, all orders must arrive with a size $\omega_x \in \left\{\pm(\sigma + k \varepsilon) | k = 0,1,2,\ldots\right\}$.  For simplicity, we assume $\sigma = \varepsilon$.} must arrive with a size $\omega_x \in \left\{\pm k \sigma | k=1,2,\ldots \right\}$.\end{mydef}

\begin{mydef}The \emph{tick size} $\pi$ of an LOB is the smallest permissible price interval between different orders within it.  All orders must arrive with a price that is specified to the accuracy of $\pi$.\end{mydef}

For example, if $\pi = \$0.00001$, then the largest permissible order price that is strictly less than $\$1.00$ is $\$0.99999$, and all orders must be submitted at a price with exactly 5 decimal places.

\begin{mydef}The lot size $\sigma$ and tick size $\pi$ of an LOB are collectively called its \emph{resolution parameters.}\end{mydef}

When a buy (respectively, sell) order $x$ is submitted, an LOB's \emph{trade-matching algorithm} checks whether it is possible to match $x$ to some other previously submitted sell (respectively, buy) order.  If so, the matching occurs immediately.  If not, $x$ becomes \emph{active}, and it remains active until either it becomes matched to another incoming sell (respectively, buy) order or it is cancelled.  Cancellation usually occurs because the owner of an order no longer wishes to offer a trade at the stated price, but rules governing a market can also lead to the cancellation of active orders.  For example, on the electronic trading platform Hotspot FX, all active orders are cancelled at 5pm EST each day to prevent an overly large accumulation of active orders over time \citep{Gould:2013statistical, HotspotGUIUserGuide}.

It is precisely the active orders in a market that make up an \emph{LOB:}

\begin{mydef}An \emph{LOB} $\mathcal{L}(t)$ is the set of all active orders in a market at time $t$.\end{mydef}

The evolution of an LOB $\mathcal{L}(t)$ is a c\`{a}dl\`{a}g process, i.e., for a limit order $x=\left(p_x,\omega_x,t_x\right)$ that becomes active upon arrival it holds that $x \in \mathcal{L}(t_x), x \notin \lim_{t'\uparrow t_x}\mathcal{L}(t')$.  The active orders in an LOB $\mathcal{L}(t)$ can be partitioned into the set of active buy orders $\mathcal{B}(t)$, for which $\omega_x<0$, and the set of active sell orders $\mathcal{A}(t)$, for which $\omega_x>0$.  An LOB can then be considered as a set of queues, each of which consists of active buy or sell orders at a specified price.

The terms \emph{bid price}, \emph{ask price}, \emph{mid price}, and \emph{bid-ask spread} are common to much of the finance literature and can be made specific in the context of an LOB:

\begin{mydef}The \emph{bid price at time $t$} is the highest stated price among active buy orders at time $t$,\begin{equation}b(t):=\max_{x \in \mathcal{B}(t)}p_x.\end{equation}The \emph{ask price at time $t$} is the lowest stated price among active sell orders at time $t$,\begin{equation}a(t):=\min_{x \in \mathcal{A}(t)}p_x.\end{equation}\end{mydef}

\begin{mydef}The \emph{bid-ask spread at time $t$} is $s(t):=a(t)-b(t)$.\end{mydef}

\begin{mydef}The \emph{mid price at time $t$} is $m(t):=\left[a(t)+b(t)\right]/2$.\end{mydef}

In an LOB, $b(t)$ is the highest price at which it is immediately possible to sell at least the lot size of the traded asset at time $t$, and $a(t)$ is the lowest price at which it is immediately possible to buy at least the lot size of the traded asset at time $t$.  It is sometimes helpful to consider prices \emph{relative} to $b(t)$ and $a(t)$.\footnote{Many different naming and sign conventions are used by different authors to describe slightly different definitions of relative price.  We introduce an explicit distinction between bid-relative price and ask-relative price to avoid potential confusion.}

\begin{mydef}For a given price $p$, the \emph{bid-relative price} is $\delta^{b}(p) := b(t)-p$ and the \emph{ask-relative price} is $\delta^{a}(p) := p-a(t)$.\end{mydef}

Observe the difference in signs between the two definitions: $\delta^{b}(p)$ measures how much smaller $p$ is than $b(t)$, and $\delta^{a}(p)$ measures how much larger $p$ is than $a(t)$.

It is often desirable to compare orders on the bid side and the ask side of an LOB.  In these cases, the concept of a single \emph{relative price of an order} is useful.

\begin{mydef}For a given order $x=(p_x,\omega_x,t_x)$, the \emph{relative price of the order} is
\begin{equation}\delta^x := \left\{ \begin{array}{ll}
\delta^b(p_x), & \text{if the order is a buy order,} \\
\delta^a(p_x), & \text{if the order is a sell order.}
\end{array}\right.\end{equation}\end{mydef}

Most traders assess the state of $\mathcal{L}(t)$ via the \emph{depth profile} or \emph{relative depth profile.}

\begin{mydef}The \emph{bid-side depth available at price $p$ and at time $t$} is\begin{equation}n^b(p,t):=\sum_{\left\{x \in \mathcal{B}(t)|p_x=p\right\}} \omega_x.\end{equation}The \emph{ask-side depth available at price $p$ and at time $t$,} denoted $n^a(p,t)$, is defined similarly using $\mathcal{A}(t)$.\end{mydef}

The depth available is often stated in multiples of the lot size.  Because $\omega_x<0$ for buy orders and $\omega_x>0$ for sell orders, it follows that $n^b(p,t)\leq0$ and $n^a(p,t)\geq0$ for all prices $p$.

\begin{mydef} The \emph{bid-side depth profile at time $t$} is the set of all ordered pairs $\left(p,n^b\left(p,t\right)\right)$.  The \emph{ask-side depth profile at time $t$} is the set of all ordered pairs $\left(p,n^a\left(p,t\right)\right)$.\end{mydef}

\begin{mydef} The \emph{mean bid-side depth available at price $p$ between times $t_1$ and $t_2$} is\begin{equation}\overline{n}^b(p,t_1,t_2) := \frac{1}{t_2-t_1}\int_{t_1}^{t_2} n^b(p,t) \mathrm{d}t.\end{equation}The \emph{mean ask-side depth available at price $p$ between times $t_1$ and $t_2$}, denoted $\overline{n}^a(p,t_1,t_2)$, is defined similarly using the ask-side depth available.\end{mydef}

Because $b(t)$ and $a(t)$ vary, it is rarely illuminating to consider the depth available at a specific price over time.  However, relative pricing provides a useful alternative.

\begin{mydef}The \emph{bid-side depth available at relative price $p$ and at time $t$} is\begin{equation}N^b(p,t):=\sum_{\left\{x \in \mathcal{B}(t)|\delta^x=p\right\}} \omega_x.\end{equation}The \emph{ask-side depth available at relative price $p$ and at time $t$,} denoted $N^a(p,t)$, is defined similarly using $\mathcal{A}(t)$.\end{mydef}

\begin{mydef} The \emph{bid-side relative depth profile at time $t$} is the set of all ordered pairs $\left(p,N^b\left(p,t\right)\right)$.  The \emph{ask-side relative depth profile at time $t$} is the set of all ordered pairs $\left(p,N^a\left(p,t\right)\right)$.\end{mydef}

\begin{mydef} The \emph{mean bid-side depth available at relative price $p$ between times $t_1$ and $t_2$} is\begin{equation}\overline{N}^b(p,t_1,t_2):=\frac{1}{t_2-t_1}\int_{t_1}^{t_2} N^b(p,t) \mathrm{d}t.\end{equation}The \emph{mean ask-side depth available at relative price $p$ between times $t_1$ and $t_2$}, denoted $\overline{N}^a(p,t_1,t_2)$, is defined similarly using the ask-side relative depth available.\end{mydef}

\begin{mydef} The \emph{mean bid-side relative depth profile between times $t_1$ and $t_2$} is the set of all ordered pairs $(p, \overline{N}^b(p,t_1,t_2))$.  The \emph{mean ask-side relative depth profile between times $t_1$ and $t_2$} is the set of all ordered pairs $(p, \overline{N}^a(p,t_1,t_2))$.\end{mydef}
Relative depth profiles provide no information about the absolute prices at which trades occur, nor do they contain information about the bid-ask spread or mid price.  However, several studies have concluded that order arrival rates depend on relative prices rather than actual prices (see, e.g., \citet{Biais:1995, Bouchaud:2002, Potters:2003, Zovko:2002}), so it is common to consider the relative depth profiles and $b(t)$ and $a(t)$ simultaneously.

\begin{figure}
\centering
\includegraphics[width=8 cm]{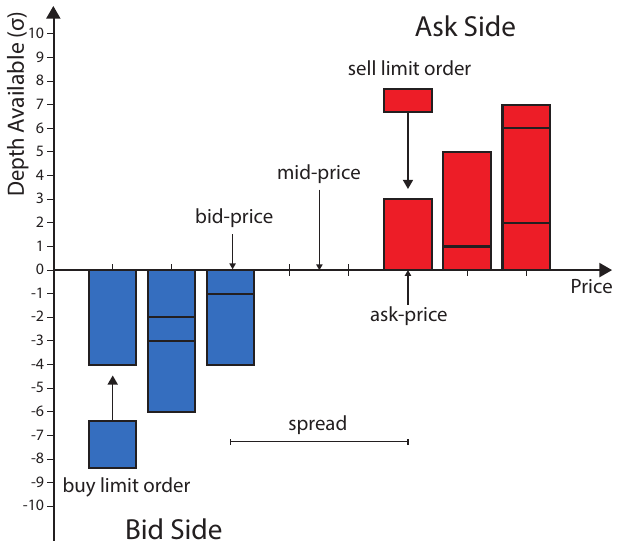}
\caption{Schematic of an LOB.}
\label{schematiclob}
\end{figure}

Figure \ref{schematiclob} shows a schematic of an LOB at some instant in time, illustrating the definitions in this section.  The horizontal lines within the blocks at each price level denote how the depth available at that price is composed of different active orders.

Time series of prices arise often during the study of LOBs.  As discussed in Section \ref{stylizedfacts}, it is a recurring theme that the behaviour of such a time series depends significantly on how it is sampled.  For example, consider the time series $m(t_1),\ldots,m(t_n)$, for some times $t_1,\ldots,t_n$.
\begin{itemize}
\item When the $t_i$ are spaced regularly in time, with $\tau$ seconds between successive samplings, such a time series is said to be sampled on a \emph{$\tau$-second timescale}.
\item When the $t_i$ are chosen to correspond to arrivals of orders, the $t_i$ may be spaced irregularly in time.  Such a time series is said to be sampled on an \emph{event-by-event timescale.}
\item When the $t_i$ are chosen to correspond to trades (i.e., matchings in an LOB), the $t_i$ may also be spaced irregularly in time.  Such a time series is said to be sampled on a \emph{trade-by-trade timescale.}
\end{itemize}

\subsection{Orders: the building blocks of an LOB}\label{theatoms}

The actions of traders in an LOB can be expressed solely in terms of the submission or cancellation of orders of the lot size.  For example, a trader who immediately sells $4\sigma$ units of the traded asset in the LOB displayed in Figure \ref{priceformation} can be considered as submitting 2 sell orders of size $\sigma$ at the price $\$1.50$, 1 sell order of size $\sigma$ at the price $\$1.49$, and 1 sell order of size $\sigma$ at the price $\$1.48$.  Similarly, a trader who posts a sell order of size $4\sigma$ at the price $\$1.55$ can be considered as submitting 4 sell orders of size $\sigma$ at a price of $\$1.55$ each.

Almost all of the published literature on LOBs adopts the following terminology.  Orders that result in an immediate matching upon submission are known as \emph{market orders}.  Orders that do not, instead becoming active orders, are known as \emph{limit orders}.\footnote{Some practitioners use the terms \emph{aggressive orders} and \emph{resting orders}, respectively, but this terminology is far less common in the published literature.}  However, it is important to recognize that this terminology is used only to emphasize whether an incoming order triggers an immediate matching or not.  There is no fundamental difference between a limit order and a market order.

Some trading platforms allow traders to specify that they wish to submit a buy (respectively, sell) market order without explicitly specifying a price.  Instead, such a trader specifies only a size, and the matching algorithm sets the price of the order appropriately to initiate the required matching.

\subsection{Price changes in LOBs}

In LOBs, the rules that govern matchings dictate how prices evolve through time.  Consider a buy (respectively, sell) order $x = (p_x, \omega_x, t_x)$ that arrives immediately after time $t$.
\begin{itemize}
\item If $p_x \leq b(t)$ (respectively, $p_x \geq a(t)$), then $x$ is a limit order that becomes active upon arrival.  It does not cause $b(t)$ or $a(t)$ to change.
\item If $b(t) < p_x < a(t)$, then $x$ is a limit order that becomes active upon arrival.  It causes $b(t)$ to increase (respectively, $a(t)$ to decrease) to $p_x$ at time $t_x$.
\item If $p_x \geq a(t)$ (respectively, $p_x \leq b(t)$), then $x$ is a market order that immediately matches to one or more active sell (respectively, buy) orders upon arrival.  Whenever such a matching occurs, it does so at the price of the active order, which is not necessarily equal to the price of the incoming order.  Whether or not such a matching causes $a(t)$ (respectively, $b(t)$) to change at time $t_x$ depends on $n^a(a(t),t)$ (respectively, $n^b(b(t),t)$) and $\omega_x$.  In particular, the new bid price $b(t_x)$ immediately after the arrival of a sell market order $x$ is\begin{equation}\label{newbid}\max(p_x,q),\text{ where }q=\arg\max_{k'}\sum_{k=k'}^{b(t)} \left| n^b(k,t)\right|>\omega_x.\end{equation}Similarly, the new ask price $a(t_x)$ immediately after the arrival of a buy market order $x$ is\begin{equation}\label{newask}\min(p_x,q),\text{ where }q=\arg\min_{k'}\sum_{k=a(t)}^{k'} n^a(k,t)>\left| \omega_x \right|.\end{equation}
\end{itemize}Put another way, the incoming order $x$ matches to the highest priority active order $y$ of opposite type.  If $\left|\omega_x\right| > \left|\omega_y\right|$, then any residue size of $x$ is considered for matching to the next highest priority active order of opposite type, and so on until either there are no further active orders with prices that make them eligible for matching, in which case the residue of $x$ becomes active at the price $p_x$, or $x$ is fully matched.  The new bid (respectively, ask) price is then equal to the price of the highest priority active buy (respectively, sell) order after the matching occurs.

\begin{figure}
\centering
\includegraphics[width=8 cm]{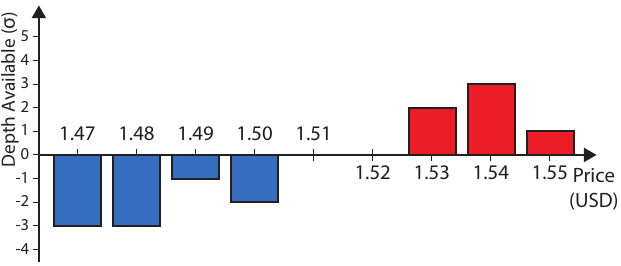}
\caption{An example LOB.}
\label{priceformation}
\end{figure}

Table \ref{priceformationtable} lists several possible market events that could occur to the LOB displayed in Figure \ref{priceformation} and the resulting values of $b(t_x), a(t_x), m(t_x),$ and $s(t_x)$ that they would cause.

\begin{table}
\begin{center}
\begin{tabular}{|c|c|c|c|c|}
\hline
 & \multicolumn{4}{|c|}{Values after arrival (USD)} \\ \cline{2-5}
Arriving order $x$ & $b(t_x)$ & $a(t_x)$ & $m(t_x)$ & $s(t_x)$ \\
\hline
Initial Values & $1.50$ & $1.53$ & $1.515$ & $0.03$ \\
$(\$1.48,-3\sigma,t_x)$ & $1.50$ & $1.53$ & $1.515$ & $0.03$ \\
$(\$1.51,-3\sigma,t_x)$ & $1.51$ & $1.53$ & $1.52$ & $0.02$ \\
$(\$1.55,-3\sigma,t_x)$ & $1.50$ & $1.54$ & $1.52$ & $0.04$ \\
$(\$1.55,-5\sigma,t_x)$ & $1.50$ & $1.55$ & $1.525$ & $0.05$ \\
$(\$1.54,4\sigma,t_x)$ & $1.50$ & $1.53$ & $1.515$ & $0.03$ \\
$(\$1.52,4\sigma,t_x)$ & $1.50$ & $1.52$ & $1.51$ & $0.02$ \\
$(\$1.47,4\sigma,t_x)$ & $1.48$ & $1.53$ & $1.505$ & $0.05$ \\
$(\$1.50,4\sigma,t_x)$ & $1.49$ & $1.50$ & $1.495$ & $0.01$ \\
\hline
\end{tabular}
\caption{How each order arrival would affect prices in the LOB displayed in Figure \ref{priceformation}.}
\label{priceformationtable}
\end{center}
\end{table}

In the financial literature, price changes are commonly studied via \emph{returns.}

\begin{mydef} The \emph{bid-price return between times $t_1$ and $t_2$} is $R^b(t_1,t_2):=(b(t_2)-b(t_1))/b(t_1)$.  The \emph{ask-price return between times $t_1$ and $t_2$,} denoted $R^a (t_1,t_2)$, and the \emph{mid-price return between times $t_1$ and $t_2$,} denoted $R^m (t_1,t_2)$, are defined similarly.\end{mydef}

\begin{mydef} The \emph{bid-price logarithmic return between times $t_1$ and $t_2$} is $r^b(t_1,t_2):=\log\left(b(t_2)/b(t_1)\right)$.  The \emph{ask-price logarithmic return between times $t_1$ and $t_2$,} denoted $r^a (t_1,t_2)$, and the \emph{mid-price logarithmic return between times $t_1$ and $t_2$,} denoted $r^m (t_1,t_2)$, are defined similarly.\end{mydef}

\subsection{The economic benefits of LOBs}

In an LOB, traders are able to choose between submitting limit orders and submitting market orders.  Limit orders stand a chance of matching at better prices than do market orders, but they also run the risk of never being matched.  Conversely, market orders never match at prices better than $b(t)$ and $a(t)$, but they do not face the inherent uncertainty associated with limit orders.  An LOB's bid-ask spread $s(t)$ can be considered as a measure of how highly the market values the immediacy and certainty associated with market orders versus the waiting and uncertainty associated with limit orders.  \citet{Foucault:2005} argued that the popularity of LOBs was due in part to their ability to allow some traders to demand immediacy, while simultaneously allowing others to supply it to those who later require it.  He conjectured that arbitrageurs, technical traders, and indexers were most likely to place market orders (due to the fast-paced nature of their work) and that portfolio managers were most likely to place limit orders (because their strategies are generally more focused on the long term).  In reality, most traders use a combination of both limit orders and market orders; they select their actions for each situation based on their individual needs at that time \citep{Anand:2005}.

\citet{Glosten:1994} argued that LOBs are an effective way for patient traders to provide liquidity to less patient traders, even when liquidity is scarce.  \citet{Luckock:2003} concluded that the volume traded in an LOB would always exceed that of a Walrasian market,\footnote{A \emph{Walrasian market} is a market in which all traders send their desired buy or sell orders to a specialist, who then determines the market value of the asset by selecting the price that maximises the volume of trade.} given the same underlying supply and demand.

\citet{Copeland:1983} noted that a limit order can be considered as a derivative contract written to the whole market, via which the order's owner offers to buy or sell the specified quantity of the asset at the specified price to any trader wishing to accept.  For example, a trader submits a sell limit order $x=\left(p_x,\omega_x,t_x\right)$ is offering the entire market a call option to buy $\omega_x$ units of the asset at price $p_x$ for as long as the order remains active.  Traders offer such derivative contracts --- i.e., submit limit orders --- in the hope that they will be able to trade at better prices than if they simply submitted market orders.  However, whether or not such a contract will be accepted by another trader (i.e., whether or not the limit order will eventually become matched) is uncertain.

\section{Challenges of Studying LOBs}\label{LOMs}

In this section, we discuss some of the challenges that LOBs present researchers.  In particular, we discuss technical issues associated with the study of empirical LOB data and present several challenges inherent in modelling LOBs.

\subsection{Perfect rationality versus zero intelligence}\label{differentperspectives}

Constructing a useful model of an LOB entails making several assumptions.  One such assumption concerns the reason that order flows exist at all.  Much of the economics literature assumes that orders are submitted because \emph{perfectly rational} traders attempt to maximize their ``utility'' by making trades in markets driven by ``information'' \citep{Parlour:2008}.  However, this assumption has come under scrutiny because utility maximisation is often inconsistent with direct observations of individual behaviour \citep{Gode:1993, Kahneman:2000choices, Lux:2009economics}.

At the other extreme lies the \emph{zero-intelligence} approach, in which aggregated order flows are assumed to be governed by specified stochastic processes whose rate parameters are conditional on other variables such as $\mathcal{L}(t)$ \citep{Daniels:2003, Cont:2010, Smith:2003}.  In this way, order flow can be regarded as a consequence of traders blindly following a set of rules without strategic considerations.  Much like perfect rationality, zero-intelligence assumptions are extreme simplifications that are inconsistent with empirical observations.  However, zero intelligence has the appeal of leading to easily quantifiable models that can yield falsifiable predictions without the need for auxiliary assumptions.  It is, therefore, a useful starting point for building models.\footnote{In Section \ref{themodel}, we explore how some authors have attempted to quantify perfect rationality for modelling purposes and discuss the often highly unrealistic assumptions that such formulations require to be tested empirically.  A detailed treatment can be found in \citet{Foucault:2005}.}

Between the two extremes of perfect rationality and zero intelligence lies a broad range of other approaches that make weaker assumptions about traders' behaviour and order flows, at the cost of resulting in models that are more difficult to study.  Many such models rely exclusively on Monte Carlo simulation to produce output.  Although such simulations still motivate interesting observations, it is often difficult to trace exactly how specific model outputs are affected by input parameters.

\subsection{State-space complexity}

It is a well-established empirical fact that current order flows depend on both $\mathcal{L}(t)$ and on recent order flows \citep{Biais:1995, Ellul:2003,Hall:2006, Hollifield:2004, Lo:2010, Sandas:2001}.  From a perfect-rationality perspective, this can be seen as traders reacting to the changing state of a market; from a zero-intelligence perspective, it can be considered as order flow rates depending on $\mathcal{L}(t)$ and on their recent history.  Either way, a key task is to uncover the structure of such conditional behaviour, either to understand what information traders evaluate when making decisions or to quantify the conditional structure of order flows.

A problem with studying conditional behaviour is that the state space of an LOB is huge: if there are $P$ different choices for price in a given LOB, then the state space of the current depth profile alone, expressed in units of the lot size $\sigma$, is $\mathbb{Z}^P$.  This makes it very difficult to investigate conditional dependences, as the number of variables is so large.  Therefore, a key modelling task is to find a way to simplify the evolving, high-dimensional state space, while retaining an LOB's important features.  Some authors have proposed ways to reduce dimensionality (see, e.g., \citet{Cont:2010c, Eliezer:1998, Smith:2003}), but there is no consensus about a simplified state space upon which very general LOB models can be constructed.

\subsection{Feedback and coupling}

In addition to traders' actions depending on $\mathcal{L}(t)$, the state of $\mathcal{L}(t)$ also clearly depends on traders' actions.  These mutual dependences induce feedback between $\mathcal{L}(t)$ and trader behaviour.  Also, as described in Section \ref{theatoms}, $b(t)$ determines the boundary condition for sell limit order placement because any sell order placed at or below $b(t)$ at least partially matches immediately.  A similar role is played by $a(t)$ for buy orders.  Therefore, order flow creates a strong coupling between $b(t)$ and $a(t)$.  \citet{Smith:2003} observed how such coupling makes LOB modelling a difficult problem.

\subsection{Priority}\label{priority}

As shown in Figure \ref{schematiclob}, several active orders can have the same price at a given time.  Much like priority is given to active orders with the best (i.e., highest buy or lowest sell) price, LOBs also employ a priority system for active orders within each individual price level.

By far the most common priority mechanism currently used is \emph{price-time.}  That is, for active buy (respectively, sell) orders, priority is given to the active orders with the highest (respectively, lowest) price, and ties are broken by selecting the active order with the earliest submission time $t_x$.  Price-time priority is an effective way to encourage traders to place limit orders \citep{Parlour:1998}.  Without a priority mechanism based on time, there is no incentive for traders to show their hand by submitting limit orders earlier than is absolutely necessary.

\begin{figure}
\centering
\includegraphics[width=8 cm]{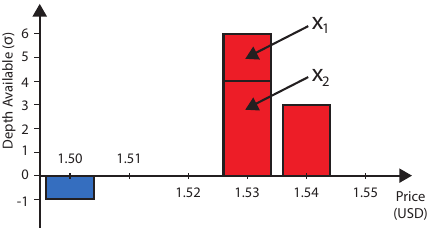}
\caption{An LOB with pro-rata priority.}
\label{proratafig}
\end{figure}

Another priority mechanism, commonly used in futures markets, is \emph{pro-rata} \citep{Field:2008}.  Under this mechanism, when a tie occurs at a given price, each relevant active order receives a share of the matching proportional to the fraction of the depth available that it represents at that price.  For example, if a buy market order of size $3\sigma$ arrived at the LOB displayed in Figure \ref{proratafig}, then $\sigma$ of it would match to active order $x_1$ and $2\sigma$ of it would match to active order $x_2$, because they correspond, respectively, to $1/3$ and $2/3$ of the depth available at $a(t)$.  Traders in pro-rata priority LOBs are faced with the substantial difficulty of optimally selecting limit order sizes, because posting limit orders with larger sizes than the quantity that is really desired for trade becomes a viable strategy to gain priority.

Another alternative priority mechanism is \emph{price-size,} in which ties are broken by selecting the active order of largest size among those at the best price.  Until recently, no major exchanges used this priority mechanism.  However, in October 2010, the first price-size trading platform, NASDAQ OMX PSX, was launched \citep{NASDAQWebsite}.  Some exchanges also allow traders to specify a minimum match size when submitting orders.  Orders with a size smaller than this are not considered for matching to such orders.  This is similar to a price-size priority mechanism: small active orders are often bypassed, effectively giving higher priority to larger orders.

Different priority mechanisms encourage traders to behave in different ways.  Price-time priority encourages traders to submit limit orders early; price-size and pro-rata priority reward traders for placing large limit orders and thus for providing greater liquidity to the market.  Traders' behaviour is closely related to the priority mechanism used, so LOB models need to take priority mechanisms into account when considering order flow.  Furthermore, priority plays a pivotal role in models that attempt to track specific orders.

\subsection{Incomplete sampling and hidden liquidity}

An LOB $\mathcal{L}(t)$ reflects only the subset of trading intentions that traders have announced up to time $t$.  However, the fact that no traders have submitted a limit order at a given price does not imply that none of them want to trade at this price, because they could be keeping their intentions private by submitting orders only when absolutely necessary \citep{Toth:2011anomalous}.  \citet{Bouchaud:2009} noted that a typical snapshot of $\mathcal{L}(t)$ at a given time is often very sparse, containing few active orders.  However, this is not an indication that few people wish to trade; it is merely an indication that they have not yet announced any intention to do so.  Indeed, some traders choose to submit only market orders and do not submit limit orders at all.\footnote{Arbitrageurs provide a key example, because their strategies depend on simultaneously buying and selling in an attempt to make instant profit.  Limit orders are of little use to them because it is uncertain when (if ever) they will be matched.}

\subsubsection{Hidden orders}

Many exchanges allow traders to conceal the extent of their intentions to trade, often at the cost of paying some penalty in terms of priority or price.  For example, many exchanges allow traders to submit \emph{iceberg orders} (also known as \emph{hidden-size orders}), a type of limit order that specifies not only a total size and price but also a \emph{visible size}.  Other traders only see the visible size.  Rules regarding the treatment of the hidden quantity vary greatly from one exchange to another.  In some cases, once a quantity of at least the visible size matches to an incoming market order, another quantity equal to the visible size becomes visible.  This quantity has priority equal to that of a standard limit order placed at this price at this time.  This sort of iceberg order is similar to a trader first submitting a limit order, then watching the market carefully and submitting a new limit order at the same price and size at the exact moment that the previous limit order matches to an incoming market order.  A trader acting in this way is sometimes deemed to be constructing a \emph{synthetic iceberg order}.  The only difference between a synthetic iceberg order and a genuine iceberg order occurs when a market order with a size larger than the (visible) depth available at the best price arrives.  In this situation, the market order matches to any visible portions of active orders at the best price according to the usual priority rules, then matches to a portion of any hidden depth available at this price.  By contrast, if a trader submits a small but entirely visible duplicate limit order immediately after the previous order matches, then a large incoming market order would match only to the active orders that existed when it arrived.  The rest of the incoming market order would instead match to the active orders at the next best price.

Some exchanges have an alternative structure for iceberg orders.  Whenever a quantity equal to at least the visible size of an iceberg order matches to an incoming market order, the rest of the order (i.e., the portion of the hidden component that does not match to the same incoming market order) is cancelled.  Iceberg orders can thereby match incoming market orders of a larger size than is initially apparent, but otherwise they behave like any other order.  This is the system currently used by the Reuters trading platform \citep{ReutersWebsite}.

Some other trading platforms allow entirely hidden limit orders.  These orders are given priority behind both entirely visible active orders at their price and the visible portion of iceberg orders at their price, but they give traders the ability to submit limit orders without revealing any information whatsoever to the market.

\subsubsection{Dark pools}

Recently, there has been an increase in the popularity of so-called \emph{dark pools} (see, e.g., \citet{Carrie:2006, Hendershott:2005}).  The matching rules governing trade in dark pools vary greatly from one exchange to another \citep{Mittal:2008}.  Some dark pools are essentially LOBs in which all active orders are entirely hidden.  Other dark pools do not allow traders to specify prices for their orders.  Instead, traders submit orders describing their desired quantity and whether they wish to buy or sell, and the dark pool holds all such requests in an entirely hidden, time-priority queue until they are matched to orders of the opposite type.  Upon matching, trades occur either at the mid-price $m(t)$ of another specified standard (i.e., non-dark) LOB for the same asset or at a price that is later negotiated by the two traders involved.

\subsubsection{Displayed liquidity}

Even in LOBs with no hidden liquidity, traders are not always able to view the set of all active orders in real time.  Many exchanges display only active orders that lie within a certain range of relative prices.  Furthermore, some electronic trading platforms only transmit updates to $\mathcal{L}(t)$ at a specific frequency, so all activity that has taken place since the most recent refresh signal is invisible to traders.

\subsection{Volatility}\label{volatility}

Loosely speaking, volatility is a measure of the variability of returns of a traded asset \citep{Barndorff:2010}.  The volatility of an asset provides some indication of how risky it is.  All else held equal, an asset with higher volatility is expected to undergo larger price changes over a given time interval than an asset with lower volatility.  For traders who wish to manage their risk exposure, volatility is an important consideration when deciding the assets in which to invest, and therefore often forms the basis of optimal portfolio construction \citep{Rebonato:2004volatility}.

Many different measures of volatility exist, and the exact form of volatility studied in a given situation depends on the type of available data and the desired purpose of the calculation \citep{Shephard:2005}.  Even when estimated on the same data, different measures of volatility sometimes exhibit different properties.  For example, different measures of volatility follow different intra-day patterns in a wide range of different markets (see \citet{Cont:2011} and references therein).  Therefore, many empirical studies report results using several different measures of volatility.
 
In an LOB, traders have access to far more information than just $b(t)$ and $a(t)$.  In particular, information such as $n^b(b(t),t)$ and $n^a(a(t),t)$ is useful to predict how prices are likely to change \citep{Biais:1995, Bortoli:2006, Ellul:2003, Hall:2006, Lo:2010}.  As discussed in Section \ref{conditionaleventfrequencies}, several empirical studies from a wide range of LOBs have reported links between volatility and other LOB properties.  However, to our knowledge, there does not yet exist an estimate of volatility that takes the full state of $\mathcal{L}(t)$ into account.  Instead, most estimates of volatility consider only changes in price series such as $b(t)$, $a(t)$, and $m(t)$.  For further discussion of practical issues regarding volatility estimation, see \citet{Liu:1999statistical}.

\subsubsection{Model-free estimates of volatility}

There is an extensive literature on the use of price series data to perform direct, model-free estimates of volatility (see, e.g., \citet{AS:2011ultra, Andersen:2010, Bandi:2006, Zhou:1996}).  In this section, we discuss three methods for performing such estimates.

\begin{mydef}Given the bid-price series $b(t_1),b(t_2),\ldots,b(t_n)$ sampled at regularly spaced times, the \emph{bid-price realized volatility} is $v^b (t_1, t_2, \ldots, t_n):=\text{st. dev.}\left(\left\{r^b (t_i,t_{i+1}) \ | \ i=1,2,\ldots,n-1 \right\}\right)$.  The \emph{ask-price realized volatility}, denoted $v^a (t_1, t_2, \ldots, t_n)$, and the \emph{mid-price realized volatility}, denoted $v^m (t_1, t_2, \ldots, t_n)$, are defined similarly.\end{mydef}

Realized volatility depends on the frequency at which price series are sampled.  It is a useful measure for comparing the variability of return series sampled with the same frequency, but it is not appropriate to compare the realized volatility of a once-daily price series for one stock to a once-hourly price series for another.

\begin{mydef}Given the bid-price series $b(t_1),b(t_2),\ldots,b(t_n)$ sampled at the times at which $n$ consecutive sell market orders arrive, the \emph{bid-price realized volatility per trade} is $V^b(t_1, t_2, \ldots, t_n):= \text{st. dev.}\left(\left\{r^b (t_i,t_{i+1}) \ | \ i=1,2,\ldots,n-1 \right\}\right)$.  The \emph{ask-price realized volatility per trade}, denoted $V^a(t_1, t_2, \ldots, t_n)$, is defined similarly using $n$ consecutive buy market order arrival times.  The \emph{mid-price realized volatility per trade}, denoted $V^m (t_1, t_2, \ldots, t_n)$, is defined similarly using $n$ consecutive market order arrival times (irrespective of whether they are buy or sell market orders).\end{mydef}

Realized volatility per trade is useful for comparing how prices vary on a trade-by-trade basis.

\begin{mydef}For a given trading day $D$, the \emph{bid-price intra-day volatility} is $\rho^b(D):=\log\left(\max_{t\in D}b(t)/\min_{t\in D}b(t)\right)$.  The \emph{ask-price intra-day volatility,} denoted $\rho^a (D)$, and the \emph{mid-price intra-day volatility,} denoted $\rho^m (D)$, are defined similarly.\end{mydef}

Intra-day volatility is useful for estimating the probability of very large price changes in a given day.  It is particularly important for \emph{day traders}, who unwind their trading positions before the end of each trading day.

\subsubsection{Model-based estimates of volatility}

A difficulty that arises when estimating any measure of volatility in an LOB is that many traders submit then immediately cancel limit orders.  This can occur for a variety of reasons, but it is often the result of electronic trading algorithms searching for hidden liquidity.  Such behaviour can cause $b(t)$ and $a(t)$ to fluctuate rapidly without any trades occurring, and it can be considered as \emph{microstructure noise} rather than a meaningful change in price.  One way to address this problem is to assume that the observed data is governed by a model from which an estimate of volatility that is absent of microstructure noise can be derived.  The parameters of the model are then estimated from the data, and the volatility estimate is derived explicitly from the model.  However, a drawback of this method is that it depends heavily on the model, and models that poorly mimic important aspects of the trading process may therefore give misleading volatility estimates.

\subsection{Resolution parameters}\label{resolutionparameters}

Values of $\sigma$ and $\pi$ vary greatly from between different trading platforms.  Expensive stocks are often traded with $\sigma=1$ share; cheaper shares are often traded with $\sigma\gg 1$ share.  In foreign exchange (FX) markets, some trading platforms use values as large as $\sigma=1$ million units of the base currency, whereas others use values as small as $\sigma = 0.01$ units of the base currency.\footnote{In FX markets, an XXX/YYY LOB matches exchanges of the \emph{base currency} XXX to the \emph{counter currency} YYY.  A price in an XXX/YYY LOB denotes how many units of currency YYY are exchanged for a single unit of currency XXX.  For example, a trade at the price $\$1.52342$ in a GBP/USD market corresponds to 1 pound sterling being exchanged for $1.52342$ US dollars.}  In equity markets, $\pi$ is often $0.01\%$ of the stock's mid price $m(t)$, rounded to the nearest power of 10.  For example, $m(t)$ for Apple Inc. fluctuated between approximately \$400 and approximately \$700 in 2012, during which time it traded with $\pi=\$0.01$.  A given currency pair is often traded with different values of $\pi$ on different trading platforms.  For example, on the electronic trading platform Hotspot FX, $\pi=\$0.00001$ for the GBP/USD LOB and $\pi=0.001$ for the USD/JPY LOB, whereas on the electronic trading platform EBS, $\pi=\$0.00005$ for the GBP/USD LOB and $\pi=0.005$ for the USD/JPY LOB \citep{EBS:2012,HotspotWebsite}.

It is a recurring theme in the literature (see, e.g., \citet{Biais:1995, Foucault:2005, Seppi:1997, Smith:2003}) that an LOB's resolution parameters $\sigma$ and $\pi$ greatly affect trade within it.  An LOB's lot size $\sigma$ dictates the smallest permissible order size, so any trader who wishes to trade in quantities smaller than $\sigma$ is unable to do so.  Furthermore, as we discuss in Section \ref{marketimpact}, traders who wish to submit large market orders often break them into smaller chunks to minimize their market impact.  The size of $\sigma$ controls the smallest permissible size of such chunks and therefore directly affects traders who implement such a strategy.  An LOB's tick size $\pi$ dictates how much more expensive it is for a trader to gain the priority (see Section \ref{priority}) associated with choosing a higher (respectively, lower) price for a buy (respectively, sell) order \citep{Parlour:2008}.  In markets where $\pi$ is extremely small, there is little reason for a trader to submit a buy (respectively, sell) limit order at a price $p$ where there are already other active orders.  Instead, he or she can gain priority over such active orders very cheaply, by choosing the price $p+\pi$ (respectively, $p-\pi$) for the limit order.  Such a setup leads to LOBs that undergo extremely frequent changes in $b(t)$ and $a(t)$ due to the small depths available.  This makes it difficult for other traders to monitor the state of the market in real time.  In September 2012, the electronic FX trading platform EBS increased the size of $\pi$ for most of its currency pairs' LOBs.  Their reason for doing so was ``to help thicken top of book price points, increase the cost of top of book price discovery and improve matching execution in terms of percent fill amounts'' \citep{EBS:2012}.

\subsection{Bilateral trade agreements}

On some exchanges, each trader maintains a \emph{block-list} of other traders with whom he or she is unwilling to trade.  A trade can only occur between traders $\theta_i$ and $\theta_j$ if $\theta_i$ does not appears on $\theta_j$'s block-list and vice-versa.  The exchange shows each trader $\theta_i$ a personalized LOB that contains only the active orders owned by traders with whom it is possible for $\theta_i$ to trade.  When a trader submits a market order, it can only match to active orders in their personalized LOB, bypassing any higher priority active orders from traders on their block-list.

On exchanges that use such \emph{bilateral trade agreements}, it is possible for a buy (respectively, sell) market order to bypass all active orders at the globally lowest (respectively, highest) price available in $\mathcal{L}(t)$ and to match to an active order with a strictly higher (respectively, lower) price.  Furthermore, given two traders $\theta_i$ and $\theta_j$ who do not have a bilateral trade agreement, it is possible for $\mathcal{L}(t)$ to simultaneously contain both an active buy order $x=(p_x,\omega_x,t_x)$ owned by $\theta_i$ and an active sell order $y=(p_y,\omega_y,t_y)$ owned by $\theta_j$, with $p_y \leq p_x$, without a matching occurring.  Therefore, it is possible for such markets to have a negative bid-ask spread.

These factors make modelling of specific matchings and of the evolution of $\mathcal{L}(t)$ a very difficult task in LOBs that operate with bilateral trade agreements.  \citep{Gould:2013statistical, Gould:2013agent} presents a full discussion of these issues, so we do not consider such LOBs further.

\subsection{Opening and closing auctions}

Many exchanges suspend standard limit order trading at the beginning and end of the trading day and instead use an auction system to match orders.  For example, the LSE's flagship order book SETS \citep{LSEWebsite} has three distinct trading phases in each trading day.  Between 08:00 and 16:30, the standard LOB mechanism is used in a period known as \emph{continuous trading}.  Between 07:50 and 08:00, a $10$-minute \emph{opening auction} takes place.  Between 16:30 and 16:35, a $5$-minute \emph{closing auction} takes place.  During both auctions, all traders can view and place orders as usual, but no orders are matched.  Due to the absence of matchings, the highest price among buy orders can exceed the lowest price among sell orders.  All orders are stored until the opening auction ends.  At this time, for each price $p$ at which there is non-zero depth available, the trade matching algorithm calculates the total volume $C_p$ of trades that could occur by matching buy orders with a price greater than or equal to $p$ to sell orders with a price less than or equal to $p$.  It then calculates the \emph{uncrossing price}\begin{equation}\hat{p}=\arg \max_p C_p.\label{uncrossingprice}\end{equation}In contrast to standard LOB trading, all trades take place at the same uncrossing price $\hat{p}$.  Given $\hat{p}$, if there is a smaller depth available for sale than there is for purchase (or vice versa), ties are broken using time priority.

Throughout the opening auction, all traders can see what the value of $\hat{p}$ would be if the auction were to end at that moment.  This allows all traders to observe the discovery of the price without any matchings taking place until the process is complete.  Such a price-discovery process is common to many markets.\footnote{\citet{Biais:1999} performed a formal hypothesis test on price-discovery data from the Paris Bourse.  Working at the 2.5\% level, they did not reject the null hypothesis that traders' conditional expectations of asset price approached the market value of the asset during the final 9 minutes of the price-discovery process.  However, they found that traders' actions were not significantly different from noise during the early part of the price-discovery process.}

\subsection{Statistical issues}

As we discuss in Section \ref{empiricals}, many authors have reported statistical regularities in LOB data from a wide variety of different markets.  However, such statistical analysis is fraught with difficulties because assumptions such as independence and stationarity, which are often required to ensure consistency of estimation, are rarely satisfied by LOB data \citep{Cont:2005,Mantegna:1999introduction}.  Furthermore, suboptimal estimators have been employed commonly in the literature, and have often produced estimates with large variance or bias.  For example, there are questions about the validity of many reported power laws throughout the scientific literature \citep{Clauset:2009, Stumpf:2012critical}.  Many authors use ordinary least-squares regression on a log-log plot to estimate power-law exponents from LOB data, yet \citet{Clauset:2009} showed that this method produces significant systematic estimation errors.  They also showed that it is inappropriate to use power-law estimators designed for continuous data on discrete data (or vice versa), yet many LOB studies do precisely this.

In this section, we list some of the pitfalls of statistical estimation using LOB data and suggest some useful estimators for data analysis.  However, these techniques are not ``one-size-fits-all'' approaches, and it is important to verify the necessary assumptions before implementing them on a given data set.

\subsubsection{Power laws}\label{powerlaws}

Several LOB properties are reported to have \emph{power-law tails:}

\begin{mydef}A random variable $Z$ is said to have a \emph{power-law tail with exponent $\alpha$} if there exists some $\alpha \in \mathbb{R}$ such that $f_Z(z)\sim O \left(z^{-\alpha}\right)$ as $z \rightarrow \infty.$\end{mydef}

If there exists a $z_{\min}>0$ such that $f_Z(z)$ is proportional to $z^{-\alpha}$ for all $z \geq z_{\min}$, then clearly $Z$ has a power-law tail.\footnote{This is not the only probability density function that has a power-law tail, but it is the most common in the literature.}  When attempting to fit power-law tails to empirical observations, it is often assumed that such a $z_{\min}$ exists (and resides within the range of the data), because the existence of such a $z_{\min}$ allows simple, closed-form expressions to be derived.  Under this assumption, \citet{Clauset:2009} provided a comprehensive algorithm for consistent estimation of $\alpha$ and $z_{\min}$ via maximum likelihood techniques, as well as for testing the hypothesis that the data really does follow a power law for $z \geq z_{\min}$.  Several other consistent estimation procedures also exist (see, e.g., \citet{Hill:1975simple, Mu:2009preferred}), but no single estimator has emerged as the best to adopt in all situations.  Therefore, some empirical studies report results using several different estimators and then draw inference based on the whole set of results.  However, as \citet{Mu:2009preferred} highlighted, different estimators often produce vastly different estimates of $\alpha$, making such inference difficult.

\subsubsection{Long-memory processes}\label{longrange}

As we discuss in Section \ref{stylizedfacts}, several time series related to LOBs have been reported to exhibit \emph{long memory}.  Intuitively, a time series has long memory if values from the present are correlated with values in the distant future.  The study of long-memory processes involves considerable challenges, and caution is needed when applying standard statistical techniques to data with long memory \citep{Beran:1994statistics}.  For example, the effective sample size of a long memory process is significantly smaller than the number of data points, so statistical estimators often converge at an extremely slow rate \citep{Farmer:2003}.  Furthermore, the correlation structure can cause such estimators to converge to arbitrary values \citep{Beran:1994statistics}.

In this section, we discuss several practical challenges of estimating long memory.  We denote by $\boldsymbol{X}$ a real-valued, second-order stationary\footnote{A time series is \emph{second-order stationary} if its first and second moments are finite and do not vary with time.  For a discussion of issues regarding stationarity in financial time series, see \citet{Taylor:2008modelling}.} time series of length $k$, $\boldsymbol{X}=X(t_1), X(t_2), \ldots, X(t_k)$.

One way to define long memory is via the asymptotic behaviour of the \emph{autocorrelation function.}

\begin{mydef}The \emph{autocorrelation function} $A$ of a time series $\boldsymbol{X}$ is given by \begin{equation}A_{\boldsymbol{X}}\left(l\right)
 :=\frac{1}{k-l}\sum_{i=1}^{k-l}\left( X(t_i)-\left\langle \boldsymbol{X} \right\rangle\right)\left(X(t_{i+l})-\left\langle \boldsymbol{X}\right\rangle\right),\end{equation}where $\left\langle \boldsymbol{X} \right\rangle = \frac{1}{k}\sum_{i=1}^k X(t_i)$ is the mean of the series.\end{mydef}

\begin{mydef}A time series $\boldsymbol{X}$ is said to exhibit \emph{long memory} if there exists some $\alpha \in (0,1)$ such that $A_{\boldsymbol{X}}$ decays like a power law,\begin{equation}\label{eq:placf}A_{\boldsymbol{X}}(l) \sim O\left(l^{-\alpha}\right), \text{ as } l \rightarrow \infty.\end{equation}
\end{mydef}

The exponent $\alpha$ describes the strength of the long memory: the smaller the value of $\alpha$, the stronger the long-range autocorrelations \citep{Lillo:2004long}.  Because of the slow decay of the autocorrelation function in a long-memory process, present values of the series can have a significant effect on its values in the distant future.  It is a recurring mistake in the literature that if $\boldsymbol{X}$ has long memory, its unconditional distribution must exhibit heavy tails.  However, \citet{Preis:2006, Preis:2007} showed that such an implication does not hold in general.

A key difficulty when using Equation (\ref{eq:placf}) to assess whether a given series has long memory is that it deals only with asymptotic behaviour.  To study the large-$l$ behaviour, it is necessary to observe more than $l$ values of $\boldsymbol{X}$, but clearly any empirically observed time series is finite.  Also, the values of the function $A_{\boldsymbol{X}}(l)$ can be small, making estimation of the functional form of $A$ very difficult.  Therefore, direct estimation of $\alpha$ from the autocorrelation function often produces very poor results \citep{Lillo:2004long}.

An alternative way to characterize long memory is via the diffusion properties \citep{Beran:1994statistics, Lillo:2004long} of the integrated series $\boldsymbol{Y}$,\begin{equation}Y(l) = \sum_{i=1}^{l}X(t_i).\end{equation}If $\boldsymbol{X}$ is a long-memory process, then the standard deviation of $\boldsymbol{Y}$ scales as $O\left(l^H\right)$, with $\frac{1}{2} < H \leq 1$ \citep{Lillo:2004long}; if $\boldsymbol{X}$ does not have long memory, then the standard deviation of $\boldsymbol{Y}$ scales as $O\left(l^{1/2}\right)$, \citep{Beran:1994statistics}.  The exponent $H$ is called the \emph{Hurst exponent,} and is related to $\alpha$ by\begin{equation}\label{hurstalpha}
H=1-\frac{\alpha}{2}.\end{equation}

Under some assumptions,\footnote{Most commonly, that $\boldsymbol{X}$ is a fractional Brownian motion, i.e., a self-similar process with Gaussian increments \citep{Beran:1994statistics, Robinson:2003time}.} there are several asymptotically unbiased estimators of $H$ that are more robust to noise in $\boldsymbol{X}$ than is direct estimation of $\alpha$ from the autocorrelation function \citep{Taqqu:1995estimators}.  However, the performance of such estimators on empirical data, which might not conform to these assumptions, varies considerably \citep{Rea:2009estimators, Xu:2005quantifying}.  Different disciplines tend to favour different estimators, although the choices are often based on historical reasons, not performance.  Some of the most commonly used are:

\begin{itemize}
\item the \emph{R/S statistic} and \emph{modified R/S statistic} \citep{Lo:1989long, Teverovsky:1999critical};
\item \emph{log-periodogram regression} \citep{Geweke:1983estimation};
\item \emph{order-$m$ detrended fluctuation analysis (DFA$m$)} \citep{Kantelhardt:2001detecting, Laspada:2011effect, Peng:1994mosaic}.
\end{itemize}

As with the estimation of power laws discussed in Section \ref{powerlaws}, no single estimator has emerged as the best in all situations, so some empirical studies report results using several different estimators and then draw inference based on them all \citep{Taqqu:1995estimators}.

\section{Empirical Observations in LOBs}\label{empiricals}

The empirical literature on LOBs is very large, yet different studies often present conflicting conclusions.  Reasons for this include different trade matching algorithms operating differently, different asset classes being traded on different exchanges, differing levels of liquidity in different markets, and different researchers having access to data of differing quality.  Furthermore, as traders' strategies have evolved over time, so too have the statistical properties of the order flow they generate.  This has become a particularly important consideration because competition and trading volumes have increased with the widespread uptake of electronic trading algorithms.

To aid comparisons, we present in Appendix \ref{empiricaltable} a description of the aims, date range, data source, and data type of each of the empirical studies of LOBs that we discuss in this survey.  We now discuss the main findings of these empirical studies in more detail, including a selection of \emph{stylized facts} that have consistently emerged from several different markets.  However, we note in Section \ref{openproblems} that there have been few recent data analyses, despite the many recent changes in markets.

\subsection{Order size}\label{ordersize}

Given the heterogeneous motivations for trading within a single market, it is unsurprising that incoming order sizes vary substantially.  Nevertheless, several regularities occur in empirical data.

For equities traded on the Paris Bourse, \citet{Bouchaud:2002} reported that the distribution of $\log(\left|\omega_x\right|)$ was approximately uniform for incoming limit orders with $\left|\omega_x\right| \in(10,50000)$.  For two stocks traded on NASDAQ, \citet{Maslov:2001} reported power-law and log-normal distributions to fit the distribution of incoming limit order sizes $\left| \omega_x \right|$.  The mean reported power-law exponent was $1\pm0.3$ (i.e., with standard deviation $0.3$).  However, the quality of the power-law fits was deemed to be weak, and the log-normal fits were deemed to be applicable over a wider range of limit order sizes than the power-law fits (although the authors stated no precise range of applicability for either).  For four stocks on the Island ECN, \citet{Challet:2001} reported that incoming limit order sizes $\left| \omega_x \right|$ clustered strongly at round-number amounts such as 10, 100, and 1000.  \citet{Mu:2009preferred} reported a similar round-number preference for market orders on the Shenzhen Stock Exchange.  \citet{Mu:2009preferred} also studied the distribution of total trade sizes when aggregated over a variety of time windows and found it to exhibit a power-law tail.  Different power-law exponent estimators produced different estimates of the tail exponent, but the authors judged the tail exponent to be larger than 2.  \citet{Maslov:2001} reported similar power-law fits on NASDAQ.  Studying 5 days of data covering 3 equities, they reported a mean power-law exponent of $1.4\pm0.1$.  Although they did not state a range of sizes over which their reported power-law distributions applied, Figure 1 in \citet{Maslov:2001} suggests an approximate range of 200 to 5000.  In a study of the 1000 largest equities in the USA, \citet{Gopikrishnan:2000} also reported power-law fits for the distribution of trade sizes.  The mean reported power-law exponent was $1.53\pm0.07$.  However, \citet{Bouchaud:2009} noted that the data studied by Gopikrishnan \emph{et al.} contained information about trades that were privately arranged to occur off-book.  They conjectured that this caused Gopikrishnan \emph{et al.} to overestimate the frequency with which very large orders occurred.

On the Stockholm Stock Exchange, \citet{Hollifield:2004} reported that buy (respectively, sell) market orders that \emph{walk up the book} --- i.e., those with a size $\left| \omega_x \right| > n(a(t),t)$ (respectively, $\omega_x > \left|n(b(t),t)\right|$) --- accounted for only $0.1\%$ of submitted market orders.  Therefore, the vast majority of submitted buy (respectively, sell) market orders matched only to active orders at $a(t)$ (respectively, $b(t)$), rather than at other prices.

\subsection{Relative price}\label{relpricearrivals}

As discussed in Section \ref{prelims}, regularities in price series are best investigated via the use of relative pricing, as $b(t)$ and $a(t)$ themselves evolve through time.  Unlike the distribution of order sizes, which seems to vary from market to market, the distribution of relative prices appears to exhibit power-law behaviour in all markets studied \citep{Bouchaud:2002, Potters:2003, Zovko:2002, Maskawa:2007correlation, Gu:2008empiricalregularities, Mike:2008}.  This may be because some traders place limit orders deep into LOBs, under the optimistic belief that large price swings could occur \citep{Bouchaud:2002}.

The distribution of relative prices of orders that arrived with non-negative relative price on the Paris Bourse \citep{Bouchaud:2002}, NASDAQ \citep{Potters:2003}, the LSE \citep{Zovko:2002, Maskawa:2007correlation}, and the Shenzhen Stock Exchange \citep{Gu:2008empiricalregularities} were reported to follow such a power law, with different values of the exponent for the different markets.  On the Paris Bourse, for buy and sell orders alike, the power-law exponent for relative prices from $\pi$ to over $100\pi$ (even up to $1000\pi$ for some stocks) was approximately $0.6$.  On NASDAQ, the ranges of relative prices over which the distributions followed a power law and the power-law exponents themselves both varied from stock to stock.  On the LSE, the value of the power-law exponent was approximately $1.5$ for relative prices between $10\pi$ and $2000\pi$ for both buy and sell orders.  In aggregated data describing all 23 studied stocks on the Shenzhen Stock Exchange, the power-law exponent for the distribution of non-negative relative prices\footnote{Observe that the notation used by \citet{Gu:2008empiricalregularities} assigns the opposite signs when measuring relative price than those that we employ.} was $1.72\pm0.03$ for buy orders and $1.15\pm0.02$ for sell orders, and the power-law exponent for the distribution of negative relative prices was $1.66\pm0.07$ for buy orders and $1.80\pm0.07$ for sell orders.  This asymmetry between buy orders and sell orders contrasts to the other studied markets, but the exact matching rules on the Shenzhen Stock Exchange prevent large price changes from occurring within a single day (which could account for this effect).

The maximum arrival rate of incoming orders was found to occur at a relative price of 0 on the LSE \citet{Mike:2008}, the Shenzhen Stock Exchange \citep{Gu:2008empiricalregularities}, the Paris Bourse \citep{Biais:1995, Bouchaud:2002}, and NASDAQ \citep{Challet:2001}, although the maximum arrival rate on the Tokyo Stock Exchange was found to occur inside the spread \citep{Cont:2010}.

\subsection{Order cancellations}

Several empirical studies covering a wide range of different markets have concluded that the vast majority of active orders end in cancellation rather than matching.  The percentage of orders that were cancelled ranged from approximately $70\%$ to more than $80\%$ on the Island ECN \citep{Hasbrouck:2002, Challet:2001}, an exchange-traded fund that tracked the NASDAQ 100 \citep{Potters:2003}, S\&P 500 futures contracts \citep{Baron:2012trading}, and in FX markets \citep{Gereben:2010brief, Lo:2010}.  Therefore, cancellations play a major role in the evolution of $\mathcal{L}(t)$ in all of these markets.

In recent years, electronic trading algorithms have surged in popularity across all markets, and such algorithms often submit and cancel vast numbers of limit orders over short periods as part of their trading strategies \citep{Harris:2002trading, Hendershott:2011}.  The widespread use of such trading algorithms seems to have further increased the percentage of orders that are cancelled in recent data.  In particular, a recent study of FX data found that more than $99.99\%$ of active orders ended in cancellation rather than matching \citep{Gould:2013statistical}.

\subsection{Mean relative depth profile}\label{mrdp}

Despite their different resolution parameters (see Section \ref{prelims}) and the different prices at which trades occur in them, several qualitative regularities are common to the mean relative depth profiles in a wide range of markets.

No significant difference was detected between the mean bid-side and the mean ask-side relative depth profiles on the Paris Bourse \citep{Biais:1995, Bouchaud:2002}, NASDAQ \citep{Potters:2003}, and Standard and Poor's Depositary Receipts (SPY)\footnote{SPY is an exchange-traded fund that allows traders to effectively buy and sell shares in all of the 500 largest stocks traded in the USA.} \citep{Potters:2003}.  By contrast, \citet{Gu:2008empiricalshape} reported asymmetry between the mean bid-side and the mean ask-side relative depth profiles on the Shenzhen Stock Exchange, but this is unsurprising considering that this market has additional rules restricting price movements each day that essentially impose asymmetric restrictions on the range of relative prices over which traders can submit orders.

Mean relative depth profiles exhibit a hump shape\footnote{More precisely, the absolute value of the mean depth available increases over the first few relative prices, and it subsequently decreases.} in a wide range of markets, including the Paris Bourse \citep{Bouchaud:2002}, NASDAQ \citep{Potters:2003}, the Stockholm Stock Exchange \citep{Hollifield:2004}, and the Shenzhen Stock Exchange \citep{Gu:2008empiricalshape}.  The maximal mean depth available for SPY was reported to occur at $b(t)$ and $a(t)$, which could also be considered as a hump with its maximum at a relative price of 0 \citep{Potters:2003}.

The location of the hump varies across markets.  However, it is difficult to perform direct comparisons between different markets because differences in their tick sizes $\pi$ affect both the granularity of the price axis and the ways in which traders behave (see Section \ref{resolutionparameters}).  There may also be strategic reasons that the hump occurs in different locations in different markets.  For example, in markets in which large price changes are relatively common, more traders can choose to submit limit orders with larger relative prices than in those where such changes are rare.  This increases the relative price at which the hump resides.  \citet{Rosu:2009} conjectured that a hump would exist in all markets in which large market orders were sufficiently likely; this represents a trade-off between the optimism that a limit order placed away from $b(t)$ or $a(t)$ might eventually be matched (at a significant profit) and the pessimism that placing limit orders too far away from $b(t)$ and $a(t)$ might never match.

\subsection{Conditional frequencies of events}\label{conditionaleventfrequencies}

The properties that we have discussed thus far in this section have all been calculated unconditionally (i.e., without reference to other events or variables).  However, several factors influence how traders interact with LOBs, so it is reasonable to study not only unconditional frequencies, but also the frequencies of those events given that some other condition was satisfied.  However, the study of such conditional event frequencies in LOBs is difficult for two main reasons:\begin{enumerate}
\item The state space is very large.  Deciding which of the enormous number of possible events or LOB states on which to condition is very difficult \citep{Parlour:2008};
\item There is a small \emph{latency} between the time that a trader sends an instruction to submit or cancel an order and the time that the exchange server receives the instruction.  Furthermore, some exchanges only send refresh signals at fixed time intervals, so traders cannot be certain that LOBs that they observe via their trading platform are perfect representations of the actual LOBs at that instant in time.  Therefore, conditioning on the ``most recent'' event is problematic, as the most recent event recorded by the exchange (and thus appearing in the market data) may not be the most recent event that a given trader observed via the trading platform.
\end{enumerate}Nevertheless, several empirical studies of conditional event frequencies in LOBs have identified interesting behaviour.  In this section, we review the key findings from several such publications, highlighting both the similarities and differences that have emerged across different markets.  Most studies have used LOB data that dates back 10 years or more.  Although this alleviates the aforementioned difficulties with latency (as the volume of order flows in LOBs was much smaller in the past than it is today, so the mean inter-arrival times between successive events were substantially longer than the latency times), it also inevitably raises the question of how representative such findings are of today's LOBs.  We return to this issue in Section \ref{openproblems}.

\subsubsection{Order size}

A simple example of conditional structure is the relationship between the size $\left|\omega_x\right|$ and the relative price $\delta^x$ of orders on the Paris Bourse, as studied by \citet{Bouchaud:2002}.  For the stocks studied, the distribution of $\left|\omega_x\right|$ varied substantially according to the relative price of the corresponding orders.  In particular, orders with larger relative price had smaller absolute size $\left|\omega_x\right|$ on average.  \citet{Maslov:2001} made a similar observation for limit orders on NASDAQ.

\subsubsection{Relative price}

On the Paris Bourse \citep{Biais:1995} and the Australian Stock Exchange \citep{Hall:2006, Cao:2008}, traders placed more orders with a relative price $\delta^x$ satisfying $-s(t) < \delta^x < 0$ (i.e., limit orders falling inside of the bid-ask spread) at times when $s(t)$ was larger than its median value.  Similarly, on the NYSE \citep{Ellul:2003}, the percentage of incoming orders that arrived with a relative price $\delta^x > -s(t)$ (i.e., were limit orders) increased as $s(t)$ increased and decreased when $s(t)$ decreased.  \citet{Biais:1995} argued that when $s(t)$ is small, it is less expensive for traders to demand immediate liquidity, so market orders become more attractive.  However, it is also possible to explain such an observation via a zero-intelligence approach: if limit order prices are chosen uniformly at random, then it is more likely that an incoming limit order price resides in the interval $(b(t),a(t))$ when the interval is wider.

On the Paris Bourse, \citet{Biais:1995} found that the percentage of buy (respectively, sell) limit orders that arrived with relative price $\delta^x$ satisfying $-s(t) < \delta^x < 0$ was higher at times when $\left|n^b(b(t),t)\right|$ (respectively, $n^a(a(t),t)$) was larger.  They conjectured that this was evidence of traders competing for priority, as the only way to gain higher priority than the active orders in the (already long) queue in this situation is to submit an order with a better price.  Furthermore, on the NYSE, \citet{Ellul:2003} reported the arrival rate of buy (respectively, sell) limit orders with a relative price $\delta^x$ satisfying $-s(t) < \delta^x < 0$ tended to increase as the total size of active buy (respectively, sell) orders increased.  They also reported a similar result for the arrival of buy (respectively, sell) market orders.  On the Australian Stock exchange, \citet{Hall:2006} calculated that the percentage of buy (respectively, sell) orders that were limit orders decreased as the total size of active buy (respectively, sell) orders increased, and \citet{Cao:2008} reported that the proportion of arriving orders that were market orders increased when $\left|n^b(b(t),t)\right|$ and $n^a(a(t),t)$ were larger.  On the LSE, \citet{Maskawa:2007correlation} concluded that traders favoured placing their limit orders at relative prices similar to those where there was already a large number of active orders.

However, such conditional structure is not common to all markets.  On the LSE, \citet{Mike:2008} reported that the distribution of relative prices was independent of $s(t)$.  On the Shenzhen Stock Exchange, \citet{Gu:2008empiricalregularities} reported that the distribution of relative prices was independent of both $s(t)$ and volatility.  On the Paris Bourse, \citet{Biais:1995} concluded that $\left|n^b(b(t),t)\right|$ (respectively, $n(a(t),t)$) had little impact on the rate of incoming sell (respectively, buy) market orders.

On the Swiss Stock Exchange, \citet{Ranaldo:2004order} reported that order flow to depend on several factors, including volatility, recent order flow, and the state of $\mathcal{L}(t)$.  Traders submitted more limit orders and fewer market orders during periods when $s(t)$ or volatility were high.  The proportion of orders that arrived with negative relative price decreased as the inter-arrival time between recent orders increased.  Traders submitted higher priced buy orders (respectively, lower priced sell orders) when the total size of active buy (respectively, sell) orders was greater.  \citet{Ranaldo:2004order} noted that buy order submission seemed to depend on both the sell side and the ask side of $\mathcal{L}(t)$, whereas sell order submission seemed to depend only on the sell side of $\mathcal{L}(t)$.  He noted, however, that market performance during the sample period might have caused such asymmetry, because the percentage change in $m(t)$ was positive for all but one of the stocks studied and exceeded $10\%$ for 4 of them.

On the LSE, \citet{Zovko:2002} reported that the relative prices of incoming limit orders were conditional on the bid-price realized volatility per trade.  They constructed two time series by calculating the mean relative price of arriving buy limit orders and the bid-price realized volatility per trade over 10 minute windows, and then calculated their cross correlation.  They rejected (at the 2.5\% level) the hypothesis that the two series were uncorrelated and observed that changes in bid-price realized volatility immediately preceded changes in mean relative price for buy limit orders.\footnote{\citet{Zovko:2002} noted that it was not clear from the cross-correlation function alone whether traders explicitly considered bid-price realized volatility when choosing the prices for their buy limit orders, or whether some other external factor first affected bid-price realized volatility and then affected traders' actions.  If the former could be demonstrated, it would support the widely-held belief that many traders consider realized volatility to be an important factor in deciding when to place a limit order \citep{Zovko:2002}.}  They also observed similar behaviour when comparing the time series of ask-price realized volatility and the time series of mean relative price for sell limit orders.

In FX markets, \citet{Lo:2010} reported that traders submitted orders with higher relative prices during periods of high mid-price realized volatility.

\subsubsection{Order flows}\label{orderflows}

On the Stockholm Stock Exchange, \citet{Sandas:2001} reported that order flows at time $t$ were conditional on both $\mathcal{L}(t)$ and on previous order flows.  In FX markets, \citet{Lo:2010} reported that order flows at time $t$ were conditional on several variables including $s(t)$, $n^b(b(t),t)$, $n^a(a(t),t)$, depth available behind the best prices, time of day, and recent order flows.  However, the precise structure of the conditional dependences varied between currency pairs.  

On the Australian Stock Exchange \citep{Hall:2006}, the arrival rates of all market events were reported to increase and decrease together.  The authors suggested that other exogenous factors (which they had not measured) might have influenced aggregate LOB activity.  In a more recent study of the Australian Stock Exchange, \citet{Cao:2008} reported that the arrival rates of market events at time $t$ were conditional on $\mathcal{L}(t)$, but not on the state of $\mathcal{L}(t)$ at earlier times.  This suggests that traders evaluated only an LOB's most recent state --- and not a longer history --- when making order placement and cancellation decisions.  \citet{Cao:2008} did not find evidence that mid-price returns had a significant impact on order arrival or cancellation rates.

Using several different financial instruments traded in electronic LOBs, \citet{Toke:2011} reported that both buy limit order and sell limit order arrival rates increased following the arrival of a market order, but they found no evidence that market order arrival rates increased following the arrival of a limit order.

\subsubsection{Event clustering}\label{eventclustering}

Using data from 40 stocks on the Paris Bourse, \citet{Biais:1995} observed strong clustering through time when studying the ``action classes'' (such as ``arrival of buy market order'', ``arrival of buy limit order within the spread'', and ``cancellation of active sell order'') of market events.  For all action classes, the conditional frequency with which a market event belonged to the specified action class, given that the previous market event also belonged to the same action class, was higher than the corresponding unconditional frequency.  The authors offered numerous possible explanations for this phenomenon: traders might have strategically split large orders into smaller chunks to avoid revealing their full trading intentions or to minimize market impact (see Section \ref{marketimpact}); different traders might have mimicked each other; different traders might have independently reacted to new information; or different traders might have tried to undercut each other (i.e., cancelled active buy (respectively, sell) orders and resubmitted them at a slightly higher (respectively, lower) price solely to gain price priority).  Bursts of small, frequent changes in $b(t)$ and $a(t)$ occurred more often when $s(t)$ was large, which they argued provided evidence of undercutting.  However, \citet{Bouchaud:2009} concluded that the phenomenon was driven primarily by strategic order splitting and found no evidence that different traders mimicked each other.

On the NYSE, \citet{Ellul:2003} reported that periods with above-average order arrival rates clustered together in time, as did periods with below-average order arrival rates.  The rate of buy (respectively, sell) limit order arrivals increased after periods of positive (respectively, negative) mid-price returns, and the rate of limit order arrivals also increased late in the trading day.  There was also a similar clustering of market events by action classes as was observed by \citet{Biais:1995} on the Paris Bourse.  However, \citet{Ellul:2003} reported that the number of occurrences of market events from a specific action class in a given $5$-minute window and the corresponding number of occurrences of market events in the previous $5$-minute window were negatively correlated.  Furthermore, they concluded that the arrival rate of market events from a given action class was more heavily conditional on the action class of the single most recent market event than it was on $\mathcal{L}(t)$, whereas the distribution of the number of occurrences of market events from a given action class in a given $5$-minute window was more heavily conditional on $\mathcal{L}(t)$ during the previous $5$-minute window than it was on the number of occurrences of market events from any specific action class in the same window.

\subsubsection{Cancellations}

On the Paris Bourse, \citet{Biais:1995} reported that cancellations of buy (respectively, sell) active orders occurred more frequently after the arrival of a buy (respectively, sell) market order.  They conjectured that this was evidence that traders submitted large orders in the hope of finding hidden liquidity and then cancelled any unmatched portions.

On the Australian Stock Exchange, \citet{Cao:2008} concluded that priority considerations played a key role for traders when deciding whether or not to cancel their active orders.  The cancellation rate for active buy (respectively, sell) orders increased when new, higher-priority buy (respectively, sell) limit orders arrived.  In addition, the cancellation rate of active buy (respectively, sell) orders at prices $p<b(t)$ (respectively, $p>a(t)$) increased when $n(p - \pi,t)$ (respectively, $n(p + \pi,t)$) became zero.  The authors proposed that this occurred because traders with active orders at price $p$ could, without substantial loss of priority, cancel and then resubmit them at price $p - \pi$ (respectively, $p + \pi$), to possibly gain a better price if the order eventually matched.  No similar increase occurred when $n(p + \pi,t)$ (respectively, $n(p - \pi,t)$) became zero.

\subsubsection{Price movements}

On the Paris Bourse, \citet{Biais:1995} reported that $a(t)$ decreased more frequently (respectively, $b(t)$ increased more frequently) immediately after the arrival of a market order that caused $b(t)$ to decrease (respectively, $a(t)$ to increase).  They suggested that such behaviour could be due to traders reacting to information, either because external sources of news caused a revaluation of the underlying asset or because traders interpreted the downward movement of $b(t)$ (respectively, upward movement of $a(t)$) itself as news.  Indeed, \citet{Potters:2003} found evidence on NASDAQ that each new trade was interpreted by traders as new information that directly affected the flow of incoming orders.

\subsubsection{Volatility}

For Canadian stocks, \citet{Hollifield:2006} reported that several different volatility measures were correlated with order flow rates.  In FX markets, \citet{Lo:2010} reported that mid-price realized volatility affected order flows.  For US equities traded on Island ECN, \citet{Hasbrouck:2002} found a lower proportion of limit orders in the arriving order flow during periods with higher volatility (using a variety of different volatility measures).  Submitted limit orders had an increased probability of execution and a shorter expected time until execution during such periods.  Furthermore, on Euronext \citep{Chakraborti:2011a} and for German Index Futures \citep{Kempf:1999}, mid-price realized volatility increased with the number of arriving market orders.  \citet{Jones:1994} reported a similar finding in a study of the NYSE; however, \citet{Ellul:2003} later reported a positive correlation between higher mid-price realized volatility and the percentage of arriving orders that were limit orders.

On the Australian Stock Exchange, \citet{Hall:2006} reported that the number of arrivals and cancellations of large limit orders (i.e., those whose size was in the upper quartile of the unconditional empirical distribution of order sizes) in any given $5$-minute window was positively correlated with mid-price realized volatility during both that window and the previous $5$-minute window.  However, in a more recent study \citep{Cao:2008} concluded that mid-price realized volatility per trade had only a minimal effect on order flows.

A weak but positive correlation between $s(t)$ and realized mid-price volatility has been observed in a wide range of markets (see \citet{Wyart:2008} and references therein).  However, a much stronger positive correlation between $s(t)$ and mid-price volatility was observed at the trade-by-trade timescale on the Paris Bourse \citep{Bouchaud:2004}, the FTSE 100 \citep{Zumbach:2004}, and the NYSE \citep{Wyart:2008}.  In a recent study of stocks traded on the NYSE, \citet{Hendershott:2011} reported that the once-daily time series of bid-price realized volatility was positively correlated with the daily mean spread.  Stocks with a lower mid price had higher bid-price realized volatilty on average.  In FX markets, \citet{Lo:2010} reported that the variance of the depth available at any given price increased during periods of high mid-price realized volatility.  \citet{Hasbrouck:2002} investigated links between volatility and various aspects of the depth profile on the Island ECN, but they found only weak relationships.

As discussed in Section \ref{differentperspectives}, \citet{Bortoli:2006} reported that mid-price intra-day volatility on the Sydney Futures Exchange varied according to how much information about the depth profile traders could view in real time.

\subsection{Market impact and price impact}\label{marketimpact}

\begin{figure}
\centering
\includegraphics[width=8 cm]{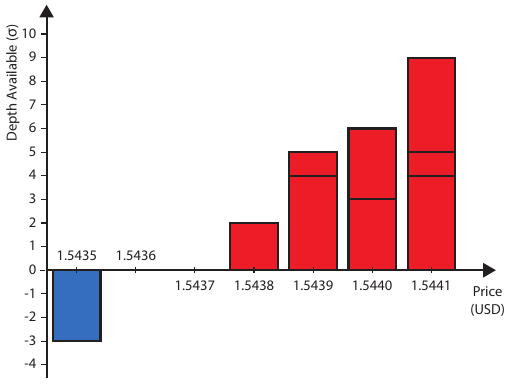}
\label{priceimpactfig}
\caption{An example LOB to illustrate price impact and market impact.}
\end{figure}

A key consideration for a trader who wishes to buy or sell a large quantity of an asset is how his or her actions might affect the asset's LOB \citep{Almgren:2001, Bouchaud:2009, Cont:2011, Eisler:2012, Obizhaeva:2013}.  For example, if trader $\theta_i$ wishes to buy $20\sigma$ shares using the LOB displayed in Figure \ref{priceimpactfig}, then submitting a single market order of size $\omega_x = -20\sigma$ would result in purchasing $2\sigma$ shares at $\$1.5438$, $5\sigma$ shares at $\$1.5439$, $6\sigma$ shares at $\$1.5440$, and $7\sigma$ shares at $\$1.5441$.  However, if $\theta_i$ were initially to submit only a market order of size $\omega_x = -2\sigma$, then it is possible that other traders might submit new limit orders, because by purchasing the $2\sigma$ shares with highest priority in the LOB, $\theta_i$ has made it more attractive for other participants to submit new sell limit orders than it was immediately before such a purchase.  If this occurs, then $\theta_i$ could submit a market order that matches to these newly submitted limit orders and then repeat this process until all $20\sigma$ shares are purchased.  Empirical observations suggest that such \emph{order splitting} is very common in a wide range of different markets \citep{Bouchaud:2009}.  Of course, there is no guarantee that the initial market order of size $2\sigma$ would stimulate such submissions of limit orders from other traders.  Indeed, it could even cause other traders to cancel their existing limit orders or to submit buy market orders, further increasing $a(t)$ and thereby ultimately causing $\theta_i$ to pay a higher price for the total purchase of $20\sigma$ shares.

The change in $b(t)$ and $a(t)$ caused by a trader's actions is called the \emph{price impact} of the actions.  The necessity for traders to monitor and control price impact predates the widespread adoption of LOBs.  In a quote-driven market, for example, any single market maker only has access to a finite inventory, so there is a limit on the size that is available for trade at the quoted prices.  Furthermore, purchasing or selling large quantities of the asset in such a market could cause market makers to adjust their quoted prices.  Both of these outcomes are examples of price impact.  In an LOB, however, it is possible to consider the impact of an action on the entire state of $\mathcal{L}(t)$.  This more general type of impact is called \emph{market impact}.  To date, the terms ``price impact'' and ``market impact'' have often been used interchangeably to refer only to changes in $b(t)$ or $a(t)$, but recent work \citep{Hautsch:2011market} has shed light on how traders' actions can affect the depths available at other prices, suggesting that it is appropriate to separate the two notions.

\citet{Bouchaud:2009} provided a detailed review of studies of both price impact and market impact.  Both are difficult to quantify formally, as they each consist of two components:
\begin{itemize}
\item \emph{instantaneous} (or \emph{immediate}) \emph{impact,} which consists of the immediate effects of a specified action;
\item \emph{permanent impact,} which consists of the long-term impact due to a specified action causing other traders to behave differently in the future.
\end{itemize} 

\noindent For example, the instantaneous price impact of a market buy order of size $2\sigma$ in the LOB in Figure \ref{priceimpactfig} is a change in $a(t)$ from $\$1.5438$ to $\$1.5439$.  An example of permanent market impact of this buy market order might be another trader deciding to submit a new sell limit order at the price $\$1.5442$.  The various forms of impact are defined as follows.

\begin{mydef} The \emph{instantaneous bid-price impact of a market event at time $t'$} is\begin{equation}b(t')-\lim_{t\uparrow t'}b(t).\end{equation}\end{mydef}

\begin{mydef} The \emph{instantaneous bid-price logarithmic return impact of a market event at time $t'$} is\begin{equation}\log b(t')-\lim_{t\uparrow t'}\left[\log b(t)\right].\end{equation}\end{mydef}

\begin{mydef} The \emph{instantaneous bid-price impact function $\phi^b (\omega_x)$} outputs the mean instantaneous bid-price impact for a buy market order of size $\omega_x$.\end{mydef}

\begin{mydef} The \emph{instantaneous bid-price logarithmic return impact function $\Phi^b (\omega_x)$} outputs the mean instantaneous bid-price logarithmic return impact for a buy market order of size $\omega_x$.\end{mydef}

Definitions for the ask price, using sell market orders of size $\omega_x$ (respectively, mid price, using both buy and sell market orders of size $\left|\omega_x\right|$) are similar.

\begin{mydef} The \emph{instantaneous market impact of a market event at time $t'$} is\begin{equation}\mathcal{L}(t') \setminus \lim_{t\uparrow t'}\mathcal{L}(t),\end{equation}where $\setminus$ denotes the difference of the two sets.\end{mydef}

Instantaneous impact exists because the arrival or cancellation of any order affects $\mathcal{L}(t)$ directly.  \citet{Bouchaud:2009} described three reasons that permanent impact might exist.  First, trades themselves might convey information to other traders.\footnote{\citet{Grossman:1980} introduced this idea for a general market, and it has since been discussed extensively in an LOB context (see, e.g., \citet{Almgren:2001, Bouchaud:2009, Hasbrouck:1991, Potters:2003}).}  Second, traders might successfully forecast short-term price movements and choose their actions accordingly.\footnote{This explanation suggests that it is not traders' actions that cause the value of an asset to rise or fall.  Instead, such changes in valuation happen exogenously and traders align their actions with them to maximize profits.  \citet{Bouchaud:2009} did not find evidence that this was a good reflection of reality.}  Third, purely random fluctuations in supply and demand might lead to permanent impact.

It is not possible to quantify precisely the permanent price or market impact of an action, because doing so would involve calculating the differences between scenarios in which the action did happen and the action did not happen.  However, all actions either happen or do not happen, so such comparisons are impossible in practice.

\subsubsection{Instantaneous price impact}

To date, instantaneous price impact for individual market orders has been studied primarily via instantaneous price impact and instantaneous logarithmic return impact functions.  On the NYSE and American Stock Exchange, \citet{Hasbrouck:1991} found $\phi^m$ to be a concave function of $\left|\omega_x\right|$.  This implies that the instantaneous price impact of a single market order of size $\left|\omega_x\right|$ was, on average, larger than the sum of the instantaneous price impacts of two market orders $x_1$ and $x_2$ of sizes $\left|\omega_{x_1}\right|$ and $\left|\omega_{x_2}\right|$, with $\left|\omega_{x_1}\right|+\left|\omega_{x_2}\right|=\left|\omega_x\right|$.

\citet{Lillo:2003} studied the stocks of 1000 different companies traded on the NYSE and sorted them into 20 groups according to their market capitalization (i.e., according to the total value of all of a given company's shares).  Within each group, they then merged their data and fitted a single curve to $\Phi^m(\left|\omega_x\right|)$.  For all 20 groups, they concluded that $\Phi^m$ followed a power law $\Phi^m(\left|\omega_x\right|) \approx \left|\omega_x\right|^{\alpha}$, with an exponent $\alpha$ that depended on the group and varied between approximately $0.2$ and $0.5$.  However, no goodness-of-fit statistics were presented with the results and it is not clear how well the fits performed for individual stocks.  After the change of variables\begin{equation}\omega_x ':=\frac{\omega_x}{C^{\eta}}, \qquad p' := pC^{\gamma},\end{equation}where $C$ was the mean market capitalization for stocks in the group and $\eta$ and $\gamma$ were fitted constants, the $\Phi^{m'}(\left|\omega_x '\right|)$ curves for each of the 20 groups collapsed onto a single curve.

\citet{Farmer:2005} reported a similar collapse of $\Phi^m$ onto a single power-law curve $\Phi^{m'}(\left|\omega_x '\right|) \approx \left|\omega_x '\right|^{0.25}$ for 11 stocks traded on the LSE after using the change of variables\begin{equation}\omega_x ' :=\frac{\omega_x\alpha}{\mu}, \qquad p' := \frac{p\lambda}{\mu},\end{equation}where $\mu$, $\lambda$, and $\nu$ denote the mean arrival rate of market orders, the mean arrival rate of limit orders, and the mean cancellation rate of active orders per unit size $\sigma$, respectively.

Using data from the Shenzhen Stock Exchange, \citet{Zhou:2012universal} partitioned incoming orders according to whether or not they received an immediate full matching\footnote{Incoming orders that are fully matched upon arrival always have a strictly smaller instantaneous mid-price impact than orders that are not.} of size $\omega_x$ at time $t_x$.  The resulting functional form of $\Phi^m(\omega_x)$ was different in the two cases.\begin{itemize}
\item For incoming orders that only partially matched upon arrival, $\Phi^m(\left|\omega_x\right|)$ was constant for all $\left|\omega_x\right|<10000$ shares, and it then increased for larger values of $\left|\omega_x\right|$.
\item For incoming orders that fully matched upon arrival, $\Phi^m(\left|\omega_x\right|)$ followed the power law $\Phi^m(\left|\omega_x\right|) \approx A\left|\omega_x\right|^{\alpha}$, where $A$ was a constant that varied across stocks.  Among buy orders, the mean value of $\alpha$ was $0.66\pm0.05$.  Among sell orders, the mean value of $\alpha$ was $0.69\pm0.06$.
\end{itemize}After applying the change of variables\begin{equation}\Phi'^m(\left|\omega_x\right|) :=\frac{\Phi^m(\left|\omega_x\right|)}{\left\langle \Phi^m \right\rangle}, \qquad \omega_x ' := \frac{\omega_x}{\left\langle \left|\omega_x \right| \right\rangle},\end{equation}where the angle brackets $\left\langle \cdot \right\rangle$ denote the mean value taken across all incoming market orders in the data, Zhou concluded that the $\Phi'^m(\omega_x ')$ curves for all studied stocks collapsed onto a single curve for incoming orders that were fully matched upon arrival and onto a different single curve for incoming orders that were only partially matched upon arrival.  The asymmetry between the bid side and the ask side was no longer present after the rescaling.

On both the Paris Bourse and NASDAQ, \citet{Potters:2003} reported that a logarithmic functional form provided a better fit to $\phi^m$ than did a power-law relationship.  Furthermore, \citet{Farmer:2003} concluded that power-law relationships overestimated the mean instantaneous mid-price impact of very large market orders on both the LSE and the NYSE.

\subsubsection{Permanent price impact}

As discussed above, it is impossible to quantify exactly the permanent price impact of a market event.  However, to gain some insight into the longer-term effects of market events, several empirical studies have compared changes in $b(t)$ and $a(t)$ over specified time intervals with measures of \emph{trade imbalance.}

\begin{mydef} The \emph{trade imbalance count during time interval $T=\left[t_1,t_2\right]$}, denoted $\Omega^c(T)$, is the difference between the total number of incoming buy market orders and the total number of incoming sell market orders that arrive during time interval $T$.\end{mydef}

\begin{mydef} The \emph{trade imbalance size during time interval $T=\left[t_1,t_2\right]$}, denoted $\Omega^\omega(T)$, is the difference between the total absolute size of all incoming buy market orders and the total size of all incoming sell market orders that arrive during time interval $T$.\end{mydef}

\citet{Evans:2002} reported a statistically significant, positive, linear relationship between the daily trade imbalance count and the ask-price logarithmic return for successive trading days in FX markets.  For German Stock Index futures, \citet{Kempf:1999} reported that the mean mid-price logarithmic return in a $5$-minute window was a concave function of the trade imbalance count during that window.  For the largest 100 stocks on the NYSE, \citet{Gabaix:2006} reported that the mean mid-price logarithmic return followed the relationship $\Omega^\omega(T)^{0.5}$ for time intervals of length $T=15$ minutes.  Using a variety of different time interval lengths for the 116 most liquid stocks in the US in 1994--1995, \citet{Plerou:2002} reported that the mean change in mid-price over the interval was a concave function of $\Omega^\omega(T)$.  Furthermore, for small values of $\Omega^\omega(T)$, the mean change in mid-price over the interval was well approximated by $\Omega^\omega(T)^{\alpha}$, where the value of $\alpha$ depended on the length of $T$.  The values of $\alpha$ ranged from $\alpha \approx 1/3$ for intervals of length 5 minutes to $\alpha \approx 1$ for intervals of length 195 minutes.  Similarly, \citet{Bouchaud:2009} reported that as the length of $T$ increased, the mean mid-price logarithmic return of the AstraZeneca stock on the LSE became better approximated by a linear function of the length $T$.

\citet{Cont:2011} recently proposed that price impact in LOBs should be studied as a function of the difference between aggregate order flow on the bid and ask sides, rather than of $\Omega^\omega(T)$.  They thereby acknowledged that cancellations can also have price impact.  Using data for 50 stockes traded on the NYSE, they performed (separately for each stock) an ordinary least-squares regression of the mean change in mid-price over a time window of length 10 seconds onto the order flow imbalance over the same time window.  For 43 of the stocks studied, the slope of the regression line was significantly different from 0 at the 95\% level and was larger on average for those stocks with smaller mean values of $\left|n(b(t),t)\right|$ and $n(a(t),t)$.  \citet{Cont:2011} noted that their ordinary least squares regressions provided a strong fit across all stocks, despite the nuances of how the individual stocks were traded.  Regressions using $\Omega^\omega(T)$ rather than order flow imbalance as the independent variable produced significantly worse fits to the data.  \citet{Cont:2011} conjectured that any observable relationship between price impact and $\Omega^\omega(T)$ was actually a byproduct of the correlation between $\Omega^\omega(T)$ and order flow imbalance.

\subsubsection{Market impact}

In contrast to the wealth of empirical studies on price impact, almost no publications address the market impact of a given action.  To our knowledge, the sole exception is the study of how order arrivals affected the state of the LOBs $\mathcal{L}(t)$ for 30 stocks on Euronext by \citet{Hautsch:2011market}.  Limit orders placed with negative relative price had a significant market impact, and limit orders placed with price $p \leq b(t)$ (respectively, $p \geq a(t)$) caused a significant permanent increase in $b(t)$ (respectively, decrease in $a(t)$).  On average, the market impact of a market order was about four times greater than that of a limit order of the same size, and limit orders with relative prices of $\pi$ or $2 \pi$ affected $b(t)$ and $a(t)$ about $20\%$ less than limit orders placed at $b(t)$ and $a(t)$.  Limit orders that arrived with non-negative relative price had no immediate market impact but significant permanent market impact.  This impact materialized more quickly for limit orders that arrived at $b(t)$ and $a(t)$ than it did for limit orders that arrived with positive relative prices.  By contrast, the market impact of limit orders placed inside of the bid-ask spread was largely instantaneous,\footnote{A buy (respectively, sell) limit order placed inside the bid-ask spread necessarily affects $b(t)$ (respectively, $a(t)$) immediately.} with little permanent impact.

\citet{Hautsch:2011market} reported similar results for all stocks studied, but they reported asymmetries between the bid side and the ask side of $\mathcal{L}(t)$, much like \citet{Kempf:1999} reported for price impact.  \citet{Hautsch:2011market} conjectured that the impact that they observed was due partly to arriving orders triggering an instantaneous imbalance in supply and demand and partly to other traders interpreting order arrivals as containing information, thereby causing them to adjust their own future actions and leading to permanent market impact.  In their view, the results suggested that the arrivals of market orders were interpreted by traders as particularly strong information signals.  This observation provides a possible explanation as to why so many traders choose to place iceberg orders: placing an iceberg order is an effective way to hide the true size of limit orders from the market and thus to minimize market impact.

\subsection{Stylized facts}\label{stylizedfacts}

Several nontrivial statistical regularities exist in empirical data from a wide range of different markets.  Such regularities are known as the \emph{stylized facts} of markets \citep{Buchanan:2011stylized}, and they might provide interesting insights into the behaviour of traders \citep{Cont:2001} and the structure of markets themselves \citep{Bouchaud:2009}.  Stylized facts are also useful from a modelling perspective, because a model's inability to reproduce one or more stylized facts can be used as an indicator for how it needs to be improved or as a reason to rule it out altogether.  For example, the existence of volatility clustering eliminates the simple random walk as a model for the temporal evolution of the mid price $m(t)$, as the existence of volatility clustering in real mid-price time series implies that large price variations are more likely to follow large price variations than they are to occur unconditionally \citep{Lo:2001}.

Reproduction of stylized facts remains a serious challenge for LOB models \citep{Chakraborti:2011a, Chakraborti:2011b, Chen:2012agent}.  This is particularly true for those based on zero-intelligence assumptions, which have tended to produce more volatile price series than empirical observations suggest is appropriate \citep{Chakraborti:2011b}.  This may imply that the strategic behaviour of real traders somehow stabilizes prices, and it is therefore an important ingredient in real LOB trading.

\citet{Cont:2001} and \citet{Chen:2012agent} both reviewed a wide range of stylized facts; we will survey a small subset that we consider to be the most relevant from an LOB perspective.  These stylized facts are of particular theoretical interest as they suggest that nonequilibrium behaviour plays an important role in LOBs.  A result from statistical mechanics is that systems in equilibrium yield distributions from the exponential family \citep{Mike:2008}, whereas distributions describing several aspects of LOB behaviour exhibit power-law tails, highlighting the possibility that LOBs might always be in a transient state.

\subsubsection{Heavy-tailed return distributions}

Over all timescales ranging from seconds to days, the unconditional distribution of mid-price returns displays tails that are heavier than a normal distribution (i.e., they have positive excess kurtosis).  Understanding heavy tails is central to risk management of investment strategies, because large price movements are more likely than they would be if returns were normally distributed.  Heavy tails have been observed on Euronext \citep{Chakraborti:2011a}, the Paris Bourse \citep{Plerou:2008stock}, the S\&P 500 index \citep{Cont:2001, Gallant:1992, Gopikrishnan:1999}, FX markets \citep{Guillaume:1997}, the NYSE \citep{Gopikrishnan:1998}, the American Stock Exchange \citep{Gopikrishnan:1998, Plerou:2008stock}, NASDAQ \citep{Gopikrishnan:1998}, the LSE \citep{Plerou:2008stock}, and the Shenzhen Stock Exchange \citep{Gu:2008empirical}.  However, the exact form of the distribution varies with the timescale used.  Across a wide range of different markets (see, e.g., \citet{Gopikrishnan:1998} and \citet{Gu:2008empirical}), the tails of the distribution at the shortest timescales appear to be well-approximated by a power law with exponent $\alpha \approx 3$, thereby earning the monicker ``the inverse cubic law of returns''.  \citet{Stanley:2008statistical} conjectured that such a universal tail might be a consequence of power-law tails in both the distribution of market order sizes and the instantaneous mid-price logarithmic return impact function.  However, \citet{Mu:2010tests} reported that this relationship did not hold in emerging markets.  \citet{Drozdz:2007stock} reported that the tails were less heavy (i.e., $\alpha>3$) in high-frequency market data for German, American, and Polish indices from 2004 to 2006, highlighting that the quantitative form of stylized facts might themselves change over time as trading styles evolve.  At longer timescales, the distribution becomes increasingly well approximated by a normal distribution.  This behaviour is often called \emph{aggregational Gaussianity} \citep{Cont:2001, Gopikrishnan:1999, Zhao:2010}.

\subsubsection{Volatility clustering}\label{longmemoryvolatility}

Time series of absolute or square mid-price returns display long memory (see Section \ref{longrange}) over timescales of weeks or even months \citep{Cont:2001, Liu:1997, Stanley:2008statistical}.  For example, the square mid-price returns for S\&P 500 index futures \citep{Cont:2001}, the NYSE \citep{Cont:2005}, the USD/JPY currency pair \citep{Cont:1997}, and crude oil futures \citep{Zhao:2010} all exhibit long memory at intra-day timescales, as do absolute mid-price returns on the Paris Bourse \citep{Chakraborti:2011a} and the Shenzhen Stock Exchange \citep{Gu:2009}.  Reported values of the Hurst exponent $H$ vary from $H \approx 0.8$ on the Paris Bourse and $H \approx 0.815$ for the USD/JPY currency pair to $H \approx 0.58$ on the Shenzhen Stock Exchange.  The long memory of absolute or square mid-price returns is often called \emph{volatility clustering} because it indicates that large price changes tend to follow other large price changes.  There are several possible explanations for volatility clustering, including the arrival of external news and the strategic splitting of orders by traders \citep{Bouchaud:2009}. 

\subsubsection{Long memory in order flow}\label{longmemoryorder}

On the LSE, \citet{Lillo:2004long} reported that the time series $n(b(t),t)$ and $n(a(t),t)$ exhibited long memory, and \citet{Zovko:2002} reported that the time series of relative prices of limit orders exhibited long memory with Hurst exponent $H \approx 0.8$.  \citet{Gu:2009} reported similar long memory in the relative prices of limit orders on the Shenzhen Stock Exchange, with $H=0.78$.  The time series constructed by assigning the value $+1$ to incoming buy orders and $-1$ to incoming sell orders has been reported to exhibit long memory on the Paris Bourse \citep{Bouchaud:2004}, the NYSE \citep{Lillo:2004long}, and the Shenzhen Stock Exchange \citep{Gu:2009}.  On the LSE, \citet{Bouchaud:2009}, \citet{Lillo:2004long}, and \citet{Mike:2008} reported that similar results hold for market order arrivals, limit order arrivals, and active order cancellations, with statistically significant differences between the estimated values of $H$ for different stocks.  However, \citet{Axioglou:2011markets} also studied the series of arriving market orders on the LSE and concluded that the apparent long memory reported by \citet{Lillo:2004long} was actually an artifact caused by market participants changing trading strategies once per day.\footnote{Stochastic processes that undergo regime switching are known to cause several estimators to report a Hurst exponent $H \neq \frac{1}{2}$ even in the absence of long memory.}

\subsubsection{Autocorrelation and long memory of returns}\label{returnseries}

Except on very short timescales, when they exhibit weak negative autocorrelation, return series do not display any significant autocorrelation \citep{Chakraborti:2011a, Cont:2005, Stanley:2008statistical}.  This well-established empirical fact has been observed in a very large number of markets, including the NYSE \citep{AS:2011ultra, Cont:2005}, Euronext \citep{Chakraborti:2011a}, FX markets \citep{Bouchaud:2003, Cont:1997}, the S\&P 500 index \citep{Bouchaud:2003, Gopikrishnan:1999}, German interest rates futures contracts \citep{Bouchaud:2003}, and crude oil futures \citep{Zhao:2010}.  The absence of autocorrelation in returns can be explained using perfect-rationality arguments \citep{Cont:2001}.  If returns were indeed autocorrelated, rational traders would employ simple strategies that used this fact to generate positive expected earnings.  Such actions would themselves reduce the level of autocorrelation, so autocorrelation would not persist.

It appears that the negative autocorrelation present on the shortest timescales disappears more quickly in more recent market data than it does in older data, which indicates that the exact quantitative details of this stylized fact may have changed over time.  Using data from the S\&P 500 index, \citet{Gopikrishnan:1999} reported negative autocorrelation in mid-price returns on timescales of up to about 20 minutes during 1984--1996 but only on timescales of up to 10 minutes during 1991--2001.  During 1991--1995, \citet{Bouchaud:2003} reported that negative autocorrelation persisted up to timescales of 20 to 30 minutes for the GBP/USD currency pair and for German interest rate futures contracts, but did not persist for timescales longer than 30 minutes.  On the NYSE, \citet{Cont:2005} reported that negative autocorrelation persisted on 5-minute timescales but not on 10-minute timescales, but did not report an exact date of when the data itself were collected.  Using data from Euronext during 2007--2008, \citet{Chakraborti:2011a} found no significant autocorrelation over time windows of 1 minute.  Furthermore, on NYSE data from 2010, \citet{Cont:2011} found no significant autocorrelation over any timescales of 20 seconds or longer.  For crude oil futures contracts traded in 2005, \citet{Zhao:2010} reported that negative autocorrelation persisted for only 10 to 15 seconds.

The various forms of long memory in order flow (see Section \ref{longmemoryorder}) might be expected to lead to long memory in return series.  However, return series on the LSE \citep{Lillo:2004long}, the Paris Bourse \citep{Bouchaud:2004}, the Deutsche Bourse \citep{Carbone:2004}, and in FX markets \citep{Gould:2013testing} all have $H \approx 0.5$ (i.e., they exhibit no long memory) on all but the shortest timescales.\footnote{There is no clear agreement about the long-memory properties of return series at the shortest timescales.  This is unsurprising, however, because microstructure effects (which vary greatly from market to market) play a prominent role in the statistical properties of return series at the shortest timescales, and estimation of $H$ is extremely sensitive to such differences in data.}  \citet{Bouchaud:2004} conjectured that this was because the long memory in price changes caused by the long memory in the arrival of market orders was negatively correlated to the long memory in price changes caused by the long memory in the arrival and cancellation of limit orders.  However, \citet{Lillo:2004long} found no evidence to support this hypothesis using data from the LSE.  Instead, they concluded that the long memory in the arrival of market orders was offset by the long memory in $n(b(t),t)$ and $n(a(t),t)$.  When predictability of market order arrivals was high, the probability that a buy (respectively, sell) market order caused a change in $m(t)$ was low, because $\left|n(b(t),t)\right|$ and $n(a(t),t)$ were large.  Therefore, the long memory in the arrival of market orders did not cause a long memory in price changes.

\section{Modelling LOBs}\label{themodel}

In recent years, the economics and physics communities have both made substantial progress with LOB modelling \citep{Chakraborti:2011b,Parlour:2008}.  However, work by the two communities has remained largely independent \citep{Farmer:2005}.  Economists have tended to be trader-centric, using perfect-rationality frameworks to derive optimal trading strategies given certain market conditions.  The LOB models produced by economists have generally treated order flow as static.  By contrast, models from physicists have tended to be conceptual toy models of the evolution of $\mathcal{L}(t)$.  By relating changes in order flow to properties of $\mathcal{L}(t)$, these models treat order flow as dynamic \citep{Farmer:2005}.  The two approaches have different strengths: an understanding of trading strategies is crucial for traders and regulators \citep{Alfonsi:2010, Almgren:2001, Cao:2008, Evans:2002, Foucault:2005, Gatheral:2010, Goettler:2006, Hall:2006, Hollifield:2006, Rosu:2010, Sandas:2001, Seppi:1997, Wyart:2008}; an understanding of the state of $\mathcal{L}(t)$ and order flow provides insight into the origins of statistical regularities, including whether they are a consequence of market microstructure or of traders' strategic behaviour \citep{Bouchaud:2009, Farmer:2005, Gu:2009, Mike:2008, Smith:2003}.

In this section, we assess existing LOB models in terms of their ability to accurately mimic the trading mechanism and to reproduce empirical facts (see Section \ref{empiricals}).  We also highlight the main modelling difficulties that are yet to be resolved.

\subsection{Perfect-rationality approaches}

In the traditional economics approach, rational investors faced with straightforward buy or sell possibilities choose portfolio strategies of holdings to maximize personal utility, subject to budget constraints \citep{Parlour:2008}.  However, LOBs provide a substantially more complicated scenario.  Rather than submitting orders for exact quantities at exact prices, an investor may attempt to construct an ideal portfolio using both limit orders and market orders.  The inherent uncertainty of execution of limit orders thereby creates uncertainty about the state of the portfolio at a given time.  When deciding whether to submit a given limit order, a trader must estimate its fill probability, which depends endogenously on both $\mathcal{L}(t)$ and future order flow.

\subsubsection{Cut-off strategies}

Many early perfect-rationality models aimed to address traders' decision-making via the use of a \emph{cut-off strategy}.

\begin{mydef}When choosing between decision $D_1$ and decision $D_2$ at time $t$, an individual employing a \emph{cut-off strategy} compares the value of a statistic $Z(t)$ with a cut-off point $z$ and makes the decision \begin{equation}\begin{array}{ll}
D_1,\quad & \text{if } Z \leq z, \\
D_2,\quad & \text{otherwise.}
\end{array}\end{equation}\end{mydef}A cut-off strategy is analogous to a hypothesis test in statistical inference.  The statistic $Z(t)$ can be any statistic related to $\mathcal{L}(t)$, current or recent order flow, the actions of other traders, and so on.  For example, a trader who wishes to place a buy order at time $t$ might decide to submit a buy market order if $s(t)$ is smaller than $5\pi$ or to submit a buy limit order otherwise.  Cut-off strategies often appear in perfect-rationality models because they drastically reduce the dimensionality of the decision space available to traders.  This is very appealing from the standpoint of tractability.

To our knowledge, the first model that addressed endogenous decision-making between limit orders and market orders in a setting that resembled an LOB was the single-period model of \citet{Chakravarty:1995}.  First, a market maker arrived and set quotes.  All other traders then arrived simultaneously and chose between submitting limit or market orders using a cut-off strategy based on the difference between their private valuations of the asset and the quotes set by the market maker.  Finally, all trades occurred simultaneously using pro-rata priority.\footnote{There is no concept of time priority in a single-period framework.}  This model demonstrated that optimal strategies for informed traders could involve submitting either limit orders or market orders, depending on how the market maker acted.  In turn, this highlighted endogenous order choice for traders as a crucial feature of a successful LOB model.  However, the inclusion of the designated market maker and the assumption that the market operated for only a single time period poorly reflects trading in real LOBs.

\citet{Foucault:1999} extended the work of \citet{Chakravarty:1995} by modelling LOB trading as a multi-step game in which traders arrived sequentially.  Limit orders remained active for only one period; if the next arriving trader did not submit a market order to match to an existing limit order, then it expired.  Upon arrival, each trader chose between placing a limit order or a market order and then left the market forever.  After each such departure, the game ended with some fixed probability; otherwise, a new trader arrived and the process was repeated.  Foucault showed that each trader's optimal strategy in this game was a cut-off strategy based on his or her private valuation of the asset and the price of the existing limit order (if one existed).

Foucault's model highlighted that the probability of matching for an active order in a LOB depends explicitly on future traders' actions (which themselves are endogenous) and that traders must actively consider other traders' strategies.  However, Foucault's model contained several assumptions that poorly mimic important aspects of real LOBs --- e.g., that limit orders remain active for only a single period and that a random, exogenous stopping time governs trading.  These assumptions restrict the model's ability to make realistic predictions about order flow dynamics and how traders estimate order fill probabilities.

\citet{Parlour:1998} studied a multi-step game in an LOB that only allowed limit orders to be submitted at a single, specific price.  Traders arrived sequentially and chose endogenously between submitting a limit order at this price or submitting a market order.  Unlike the model from \citet{Foucault:1999}, limit orders did not expire.  The work identified explicit links between traders' strategies and $\mathcal{L}(t)$.  In particular, Parlour demonstrated that the optimal decision between submitting a limit order or a market order should be made by employing a cut-off strategy that assessed both sides of $\mathcal{L}(t)$ to estimate the fill probability for a limit order.  If the estimated fill probability was sufficiently high, then the trader should submit a limit order; otherwise, he or she should submit a market order.  Parlour argued that limit orders become less attractive later in a trading day due to their lower fill probabilities before the end of trading.  However, by disallowing cancellations of active orders and by restricting the pricing grid to a single value, Parlour's model simplified the decision-making process of how to act in real LOBs \citep{Hollifield:2006}.

\citet{Hollifield:2004} tested the hypothesis that cut-off strategies such as those discussed above could explain the observed actions of traders trading the Ericsson stock on the Stockholm Stock Exchange.  Working at the 1\% level, they accepted this hypothesis when the bid side or the ask side of $\mathcal{L}(t)$ were each considered in isolation but rejected it when both sides of $\mathcal{L}(t)$ were considered together, due to the existence of several limit orders with extremely low fill probabilities whose expected payoffs were too low for the model to justify.  \citet{Hollifield:2004} concluded that cancellations, which were absent from the models discussed above, must play an important role in real LOBs.

\citet{Hollifield:2006} studied a model in which cancellations were endogenous.  By comparing predictions made by the model to data from the Vancouver Stock Exchange, they concluded that real traders did not make decisions using a common cut-off strategy.

\subsubsection{Fundamental values and informed traders}

Some perfect-rationality models centre around the idea that a subset of traders are \emph{informed traders} who know the ``fundamental'' or ``true'' value of the traded asset, whereas everyone else is \emph{uninformed} and does not know this true value (see, e.g., \citet{Copeland:1983, Glosten:1985, Glosten:1994, Kyle:1985}).  \citet{Bouchaud:2009} noted that many researchers now reject the idea that assets have fundamental values, but such models can still provide insight into price formation in markets with asymmetric information.

In the classic \citet{Kyle:1985} model, uninformed traders placed limit orders and market orders in an LOB.  At the same time, informed traders observed this LOB and, if an uninformed trader posted a buy limit order with a price above (respectively, sell limit order with a price below) the fundamental value, then an informed trader submitted a market order that matched to the mispriced limit order and thereby made a profit.  However, more recent models \citep{Chakravarty:1995, Rosu:2010} have noted several reasons that informed traders should sometimes choose to submit limit orders rather than market orders --- for example, to avoid detection by other traders who would mimic their actions if they knew them to be informed \citep{Rosu:2010}.

\citet{Goettler:2006} studied a model in which traders arrived at an LOB following a Poisson process.  Upon arrival, a trader submitted any desired orders, choosing freely among prices.  He or she then left the market and re-arrived following an independent Poisson process.  Upon re-arrival, a trader could cancel or modify his or her active orders.  When a trader performed a trade, he or she left the market forever.  Additionally, any trader could, at any time, pay a fee to become informed about the fundamental value of the asset.  Such traders stayed informed until they eventually traded and left the market.  Goettler \emph{et al.} concluded that a trader's willingness to purchase the information should decrease as his or her desire to trade increases.  They concluded that speculators, who trade purely for profit, should buy the information most often, that the value of the information increased with volatility, and that the optimal strategy for an informed trader included submissions of both limit orders and market orders.  However, as \citet{Parlour:2008} discussed, Goettler \emph{et al.'s} step forward in realism came at the cost of discarding analytical tractability, which forced them to rely solely on numerical computations.

\citet{Rosu:2010} also investigated how informed traders should choose between limit orders and market orders in a model that allowed cancellations.  He showed that if an informed trader observed a mispricing above a given threshold, he or she should submit a market order to capitalize on the opportunity before anyone else.  If the mispricing was below this threshold (but still positive), the trader should instead a limit order, to gain a better price for the trade if it matched.  Ro\c{s}u also concluded that the price impact of a single informed trader's order submissions were insufficient to reset $b(t)$ and $a(t)$ to their fundamental levels, so subsequent informed traders who arrived at the market with the same information were able to perform similar actions to make a profit.  He argued that this provided a possible explanation for the empirically observed phenomenon of event clustering (see Section \ref{orderflows}).

\citet{Rosu:2009} replaced the idea that traders who selected different prices for their orders must have done so because of asymmetric information \citep{Glosten:1985, Kyle:1985} with the notion that different traders might have selected different prices for their orders because they valued the immediacy of trading differently.  For example, in real markets, some traders need to trade immediately and therefore submit market orders; others do not and can therefore submit limit orders in the hope of eventually trading at a better price.  In Ro\c{s}u's model, traders were able to modify and cancel their active orders in real time, making it the first perfect-rationality LOB model to reflect the full range of actions available.  Rather than complicating the model, Ro\c{s}u demonstrated that limit order cancellations simplified the decision-making problem.  He proved the existence of a unique Markov-perfect equilibrium in the game and derived the optimal strategy for a newly arriving trader.  He also showed that a hump-shaped depth profile emerged in an LOB populated by traders following such a strategy, in agreement with empirical findings from several different markets (see Section \ref{mrdp}).

\subsubsection{Minimizing market impact}

As discussed in Section \ref{marketimpact}, determining how to minimize the market impact of an order is a key consideration for traders.  Several perfect-rationality models have suggested that the event clustering found in empirical data (see Section \ref{orderflows}) may be a signature of traders attempting to minimize their market impact when executing large orders \citep{Bouchaud:2009}.  \citet{Lillo:2005} showed that the power-law decaying autocorrelation function exhibited by order flows present in empirical data could be reproduced by a model in which traders who wished to buy or sell large quantities of an asset did so by submitting a collection of smaller orders sequentially through time.

Using a discrete-time framework, \citet{Bertsimas:1998} derived an optimal trading strategy for a trader who sought to minimize expected trading costs, including those due to market impact, when processing a very large order that had to be completed in the next $k$ time steps.  They showed that if prices followed an arithmetic random walk, then the trader should split the original order into $k$ equal blocks, and submit the blocks uniformly through time.  They also showed that if prices reflected some form of exogenous information, then the optimal strategy involved dynamically adjusting trade quantities at each step.  \citet{Almgren:2001} derived a similar strategy for traders who maximized the utility of trading revenues, including a penalty for uncertainty, when executing a large order.  However, both of the assumptions about the behaviour of prices in these models poorly mimic the structure of empirically observed price series \citep{Lo:2001}.  

\citet{Obizhaeva:2013} considered the \emph{optimal execution} problem in continuous time.  In this setup, it is also necessary to choose optimal times, not just optimal sizes, at which to submit orders.  The authors showed that considering the limit $k \rightarrow \infty$ of a $k$-period discrete-time model did not provide a valid solution to the problem, as it led to a degenerate case where execution costs were strategy-independent.  By making several strong assumptions --- including that the depth profile underwent exponential recovery in time\footnote{Discussion about such recovery of the depth profile, often known as its \emph{resiliency}, has appeared in both the empirical \citep{Biais:1995, Bouchaud:2004, Potters:2003} and modelling literatures \citep{Foucault:2005, Rosu:2009}.} back to a neutral uniform state of $n(p,t)=n(p',t)$ for all prices $p,p'$ after the arrival of a market order, Obizhaeva and Wang derived explicit optimal execution strategies and concluded that the theoretical optimum required the submission of uncountably many orders during a finite time period.  \citet{Alfonsi:2010} developed the model further by removing the assumption that the neutral state of the depth profile must be uniform, although they still assumed that recovery to the neutral state was exponential.  They showed that in continuous time, the optimal execution strategy involved initially submitting a large market order to stimulate new limit order submissions and then submitting small, equally-sized market orders at a fixed rate, then submitting another large market order at the end.

\subsection{Zero-intelligence approaches}

As noted above, most perfect-rationality models rely on a series of auxiliary assumptions to quantify unobservable parameters.  Such assumptions often make it difficult to relate perfect-rationality models real LOBs.  By contrast, zero-intelligence models assume that order arrivals and cancellations are directly governed by stochastic processes.  The parameters of such stochastic processes can be estimated directly from historical data, and the statistical properties of the models' outputs can be compared to those of real data.  In this way, falsifiable hypotheses can be formulated and tested empirically.  Furthermore, the predictive power of models can be measured by training them on a subset of available data in-sample and then evaluating them against other data out-of-sample.

\subsubsection{Model framework}

\begin{figure}
\centering
\includegraphics[width=0.45\textwidth]{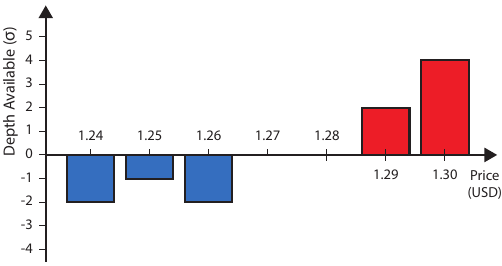}
\includegraphics[width=0.45\textwidth]{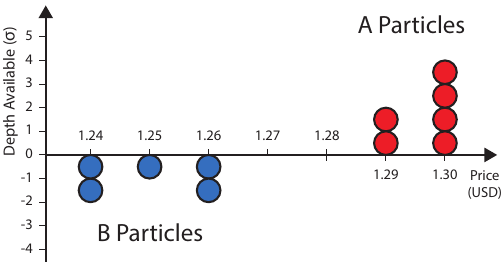}
\caption{An LOB and its corresponding representation as a system of particles on a one-dimensional pricing lattice.}
\label{particles}
\end{figure}

Most zero-intelligence LOB models use the framework introduced by \citet{Bak:1997} to model the evolution of $\mathcal{L}(t)$.  Orders are modelled as particles on a one-dimensional lattice whose locations correspond to price.  Sell orders are represented as a particle of type $A$ and buy orders are represented as a particle of type $B$ (see Figure \ref{particles}).  Each particle corresponds to an order of size $\sigma$, so an order of size $k\sigma$ is represented by $k$ separate particles.  When two orders of opposite type occupy the same point on the pricing grid, an annihilation $A+B\rightarrow \emptyset$ occurs.

\subsubsection{Random-walk diffusion models}

\citet{Bak:1997} introduced the earliest class of zero-intelligence LOB models involving particles diffusing along a price lattice.  Given an initial LOB state with all $A$ particles to the right of all $B$ particles, they modelled the movement of each particle along the price lattice using a random walk.  Several authors studied such models analytically and via Monte Carlo simulation \citep{Bak:1997, Chan:2001, Eliezer:1998, Tang:1999}.  Such work produced several possible explanations for empirical regularities observed in real LOB data, such as the hump-shaped depth profile (see Section \ref{mrdp}).  However, the \citet{Bak:1997} model has since been rejected because the diffusion of active orders across different prices is not observed in LOBs \citep{Chakraborti:2011b, Challet:2001, Farmer:2005}.  Nonetheless, these models sparked the idea that empirical regularities in LOB data that were previously thought to be a direct consequence of traders' strategic actions could be reproduced in a zero-intelligence framework.  This has subsequently become a central theme of zero-intelligence models throughout the literature (see, e.g., \citet{Bouchaud:2009, Farmer:2005, Farmer:2009, Smith:2003}).

\subsubsection{Discrete-time models}

\citet{Maslov:2000} introduced a model that bore a stronger resemblance to a real LOB than the price diffusion models discussed above.  In Maslov's model, a single trader arrived at each discrete time step.  With probability $1/2$, this trader was a buyer; otherwise, he or she was a seller.  Independently, with probability $1-r$ the trader submitted a market order; otherwise, he or she submitted a limit order $x=(p_x, \sigma, t_x)$ with\begin{equation}p_x=\left\{\begin{array}{ll}
p'- K, & \text{ if the trader was a buyer,} \\
p'+ K, & \text{ if the trader was a seller,}
\end{array}\right.\end{equation}where $p'$ was the most recent price at which a matching had occurred and $K$ was a random variable with a specified distribution.  No cancellations or modifications to active orders were allowed.  Even with only $1000$ iterations and in very simple setups (such as $r=1/2$ and $K=1$ with probability 1; or $r=1/2$ and $K \sim \mathrm{Uniform}\left\{1,2,3,4\right\}$), the return series generated by the model was found to exhibit heavy-tails and negative autocorrelation at low lags on event-by-event timescales.  \citet{Slanina:2001} showed that the heavy-tailed return distribution and negative autocorrelation remained when implementing a mean-field approximation to replace tracking the prices of individual limit orders with that of a mean value that increased when a limit order arrived and decreased when a market order arrived.  However, this model generated mid-price returns with a Hurst exponent of $H \approx 0.25$ on all timescales.  By contrast, as discussed in Section \ref{stylizedfacts}, LOB data displays no long memory (i.e., $H \approx 0.5$) in mid-price returns on all but the shortest timescales \citep{Lillo:2004long}.

\citet{Challet:2001} refined Maslov's model by allowing multiple particles to be deposited on the pricing grid during a single timestep.  They also allowed existing particles to evaporate, corresponding to the cancellation of an active order, although such evaporations were assumed to occur exogenously and independently for each particle.  Challet and Stinchcombe's model exhibited a heavy-tailed return distribution and volatility clustering, and the Hurst exponent of the mid-price return series at large timescales was $H \approx 0.5$.  The authors conjectured that the evaporations in their model (which were absent in the model of \citet{Maslov:2000}) ensured that the Hurst exponent at large timescales matched that of empirical data.

\subsubsection{Continuous-time models}

The first zero-intelligence model in continuous time was introduced by \citet{Daniels:2003}, who produced a master equation for $\mathcal{L}(t)$ under tha assumptions that market order arrivals, limit order arrivals, and cancellations were all governed by independent Poisson processes, and that incoming limit orders arrived at the same rate at each relative price in the semi-infinite interval $(-s(t),\infty)$.  \citet{Smith:2003} solved the master equation using a mean-field approximation that the depths available at neighbouring prices were independent, in the limit of infinitesimal tick size $\pi \rightarrow 0$.  Guided by dimensional analysis, they constructed simple, closed-form estimators for a variety of LOB properties, such as the mean spread, mean depth available at a given price, and mid-price diffusion, in terms of only the lot size $\sigma$ and the Poisson processes' arrival rates.  Monte Carlo simulations produced similar results.  The model also provided possible explanations for why some empirical properties of LOBs varied between different markets (see Section \ref{empiricals}).  In particular, the lot size $\sigma$ appeared explicitly in many of the closed-form estimators, and there were phase transitions between different types of market behaviour as $\sigma$ varied.

Many of the assumptions made by \citet{Daniels:2003} and \citet{Smith:2003} to maintain analytical tractability provide poor resemblance to some aspects of real LOBs.  For example, in the limit $\pi \rightarrow 0$, the only possible numbers of limit orders that can reside at a given price $p$ are 0 and 1.  This destroys the notion of limit orders queueing up at given prices and thereby removes a primary consideration for traders: when to submit an order at the back of an existing priority queue versus when to start a new queue at a worse price (see Section \ref{resolutionparameters}).  Despite its simplifications, \citet{Farmer:2005} reported that the model performed well when tested against some aspects of empirical data.  In particular, they made predictions of the mean spread and a measure of price diffusion\footnote{\citet{Farmer:2005} studied price diffusion by calculating the variance $v_\tau$ of the set $\left\{(m(t_i+\tau) - m(t_i)) \ | \ i=1,\ldots,n\right\}$ for various values of $\tau$, where $\left\{t_i \ | \ i=1,\ldots,n\right\}$ is the set of times at which the mid-price changed.  They then performed an ordinary least-squares regression to estimate $d$ in the expression $v_{\tau}=d\tau$.} for 11 stocks traded on the LSE by calibrating the model's parameters using historical data and then compared these predictions to the real data using an ordinary least-squares regression:\begin{equation}Z_{\text{emp}}(i)=zZ_{\text{mod}}(i)+c,\end{equation}where $Z_{\text{emp}}(i)$ and $Z_{\text{mod}}(i)$ are the mean empirical and model output values of statistic $Z$ for stock $i$.  Using this setup, $z=1$ and $c=0$ correspond to a perfect fit of the model to the data.  For the mean spread, the ordinary least-squares estimates of the parameters were $z \approx 0.99\pm0.10$ and $c \approx 0.06\pm0.26$.  For the price diffusion, the ordinary least-squares estimates of the parameters were $z=1.33\pm0.10$ and $c=0.06\pm0.26$.  \citet{Farmer:2005} used bootstrap resampling to estimate the standard errors of the regression coefficients, because serial correlations within the data invalidate the assumptions required to use the standard estimators (see Section \ref{longrange}).  However, the distribution of mid-price returns did not display heavy tails, and \citet{Toth:2011anomalous} reported that time series of logarithmic mid-price returns generated by the model had a Hurst exponent $H < \frac{1}{2}$ when the model's parameters were chosen to mimic realistic market conditions.  Both of these facts are contrary to findings in empirical data (see Section \ref{stylizedfacts}).

\citet{Cont:2010} recently introduced a variant of the \citet{Daniels:2003} and \citet{Smith:2003} model, to understand how the occurrence frequency of certain events was conditional on $\mathcal{L}(t)$.  The model did not assume that $\pi \rightarrow 0$, and thereby ensured that priority queues formed at discrete points on the price lattice.  \citet{Cont:2010} also removed the assumption of \citet{Daniels:2003} and \citet{Smith:2003} that the relative prices of limit orders were drawn from a uniform distribution, and replaced it with a power-law distribution to fit better observations from empirical data \citep{Bouchaud:2002, Cont:2010, Potters:2003, Zovko:2002}.  Simulations of the \citet{Cont:2010} model exhibited the hump-shaped depth profile commonly reported in empirical data (see Section \ref{empiricals}).  Using Laplace transforms, the authors computed conditional probability distributions for the matching of limit orders in given situations.

\citet{Zhao:2010} and \citet{Toke:2011} recently extended the \citet{Cont:2010} model by revising the assumed arrival structure of market events.  Based on an empirical study of crude oil futures traded at the International Petroleum Exchange, \citet{Zhao:2010} rejected the assumption that the inter-arrival times of market events were independent draws from an exponential distribution and thereby rejected the use of independent Poisson processes to model market event arrivals.  Zhao replaced the independent Poisson processes with a Hawkes process\footnote{A \emph{Hawkes process} is a point process with time-varying intensity parameter $\lambda(t)=\lambda_0(t)+\sum_{t_i < t} \sum_{j} C_j e^{-D_j(t-t_i)}$, where $t_i$ denotes the time of the $i^{\text{th}}$ previous arrivals and $C_j$ and $D_j$ are parameters that control the intensity of arrivals.} \citep{Bauwens:2009} that described the arrival rate of all market events as a function of recent order arrival rates and of the number of recent order arrivals.  When an arrival occurred, its type (e.g., market order arrival, limit order cancellation, etc.) was determined exogenously.  This produced order flows in which periods of high arrival rates clustered together in time and in which periods of low arrival rates clustered together in time, in agreement with empirical data \citep{Ellul:2003, Hall:2006}.  Zhao demonstrated that this improved the fit of the model output to the empirically observed mean relative depth profile.  \citet{Toke:2011} similarly replaced the Poisson processes in the \citet{Cont:2010} model with Hawkes processes.  Unlike Zhao, however, Toke used multiple mutually-exciting Hawkes processes (one for each type of market event).  By studying empirical data from several different asset classes, Toke observed that when a market order arrived, the mean time until the next limit order arrival was less than the corresponding unconditional mean time.  The Hawkes processes produced simulated order flow and spread dynamics that matched empirical observations more closely than those produced by a Poisson process model.

\citet{Cont:2010c} recently introduced a model that tracked only $n^b(b(t),t)$ and $n^a(a(t),t)$ rather than the whole depth profile.  When either became zero, the model assumed that the depth available at the next best price was a random variable drawn from a distribution $f$.  The state space of this model was $\mathbb{N}^{2}$, rather than $\mathbb{Z}^P$ as in most other recent LOB models.  The authors' justification for such a simplified setup was that many traders can only view the depths available at the best prices, and not the entire depth profile (although this is increasingly less common as electronic trading platforms deliver ever more up-to-date information in real time \citep{Boehmer:2005, Bortoli:2006}).  Independent Poisson processes governed market order arrivals, limit order arrivals, and limit order cancellations.  Using only the Poisson processes' rate parameters and the distribution $f$, the authors derived analytical estimates for several market properties --- including volatility, the distribution of time until the next change in $m(t)$, the distribution and autocorrelation of price changes, and the conditional probability that $m(t)$ moved in a specified direction given $n^b(b(t),t)$ and $n^a(a(t),t)$.  Different levels of autocorrelation of the mid-price series emerged at different sampling frequencies, in agreement with empirical observations \citep{Cont:2001, Zhou:1996}.

\subsubsection{Beyond zero intelligence}

\citet{Toth:2011anomalous} extended the \citet{Daniels:2003} and \citet{Smith:2003} model by using a long-memory process to determine whether arriving orders were buy or sell orders.  They also introduced an extra parameter to relate the size of arriving buy (respectively, sell) market orders to $n^a(a(t),t)$ (respectively, $n^b(b(t),t)$).  This extra parameter made it possible to control the strength of long memory in the logarithmic mid-price return series generated by the model, thereby addressing an issue with the original model.

Based on an empirical study of data from the LSE, \citet{Mike:2008} incorporated the empirically observed long memory of order flow (see Section \ref{longmemoryorder}) into their model of the evolution of $\mathcal{L}(t)$.  They used a Student's $t$ distribution to model the relative prices of incoming orders, and they closely matched cancellation rates for active orders to empirical data.  For stocks with small tick size and low volatility, they found that their model exhibited negative autocorrelation of logarithmic mid-price returns on short timescales, in agreement with empirical data.  Furthermore, they found that it made good predictions of the distribution of mid-price returns (including heavy tails) and of $s(t)$.  However, for stocks other than those with small tick size and low volatility, it was less successful.

\citet{Gu:2009} simulated the \citet{Mike:2008} model and performed a DFA$m$ (see Section \ref{longrange}) on the output mid-price return and volatility series.  They found that the mid-price return series did not exhibit long memory, in agreement with empirical data, but that neither did the volatility series.  This is in disagreement with the widely observed stylized fact of volatility clustering (see Section \ref{stylizedfacts}).  Gu and Zhou then proposed an extension to the model in which the relative prices of orders were not drawn independently, but rather were simulated to exhibit long memory.  This modification caused long memory to emerge in the volatility series and preserved all of the model's other results.

\citet{Gu:2009probability} replaced several of the stochastic processes governing order flow in the \citet{Mike:2008} model with other distributions to examine how this affected the output.  They concluded that a power-law tail in the mid-price return distribution only appeared in the model's output when the distribution from which positive relative prices were drawn had heavy tails, irrespective of whether or not the distribution from which negative relative prices were drawn had heavy tails.

Although \citet{Toth:2011anomalous} and \citet{Mike:2008} did not directly assume that traders were rational, the conditional structure of random variables in their models can clearly be considered as consequences of rational decision-making.  For example, the dependence of market order sizes on $n^a(a(t),t)$ and $n^b(b(t),t)$ in the \citet{Toth:2011anomalous} model can be interpreted as traders attempting to minimize their market impact, and the lower rate of cancellation among active orders with larger relative prices in the \citet{Mike:2008} model can be considered to reflect how traders would not submit such orders unless they were willing to wait for them to be filled in the future.

\subsection{Agent-based models}

An \emph{agent-based model (ABM)} is a model in which a large number of possibly heterogeneous agents interact in a specified way \citep{Gilbert:2007agent}.  A key advantage of ABMs is the ability to incorporate heterogeneity between different traders \citep{Buchanan:2008little, Chakraborti:2011b}.  Such models can provide insight into both the performance of individual agents and the aggregate effect of all agents' interactions.  By allowing each individual agent's behaviour to be specified without any explicit requirements regarding rationality, ABMs lie between the two extremes of zero-intelligence and perfect-rationality models.  However, ABMs of LOBs also have significant drawbacks.  Due to the large number of interacting components in an LOB, it is difficult to track explicitly how a specified input parameter affects the output of the ABM.  It is also very difficult to encode a quantitative set of rules that describes traders' complex and interacting strategies, and finding a set of agent rules that produces a specific behaviour from an ABM provides no guarantee that such a set of rules is the only one to do so \citep{Preis:2007}.  \citet{Abergel:2011} attempted to address these issues by studying systems of stochastic differential equations describing price dynamics in terms of some ABMs' input parameters, thereby deriving exact links between the two approaches.  For example, they demonstrated that a very simple ABM resulted in Gaussian process dynamics, with a diffusion coefficient that depended on the model's input parameters.

Early ABMs of LOBs assumed that agents arrived sequentially \citep{Foucault:1999} and that LOBs emptied at the end of each time step.  Such setups fail to acknowledge an LOB's key function of storing supply and demand for later consumption by other traders \citep{Smith:2003}.  However, more recent ABMs have more closely mimicked real LOBs and have successfully reproduced a wide range of empirical features present in empirical data \citep{Challet:2003, Chiarella:2002, Cont:2000, Preis:2006}.

\citet{Cont:2000} showed that when agents in a simple market imitated each other, the resulting output exhibited a heavy-tailed return distribution, clustered volatility, and aggregational Gaussianity (see Section \ref{stylizedfacts}).

\citet{Chiarella:2002} studied an ABM in which all agents shared a common valuation for the asset traded in a given LOB.  They noted that the realized volatility produced by their model was too low compared to empirical data and that there was no volatility clustering.  They thereby argued that substantial heterogeneity must exist between traders in real LOBs in order for the highly nontrivial properties of volatility to emerge (see Section \ref{volatility}).  \citet{Cont:2005} noted that differences in agents' levels of impatience could be a source of such heterogeneity in real markets.

\citet{Preis:2006} reproduced the main findings of \citet{Smith:2003} using an ABM rather than independent Poisson processes.  By fine-tuning agents' trading strategies, the model reproduced the heavy-tailed distribution of mid-price returns, the diffusivity of mid-price returns over long timescales, and the negative autocorrelation of $m(t)$ on an event-by-event timescale.  \citet{Preis:2007} also studied the performance of individual agents in the model.  He found that the Hurst exponent $H$ of the mid-price return series depended on the number of agents in the model, and that the best fit of $H$ against values calculated from empirical data occurred with 150 to 500 liquidity provider (i.e., limit order placing) agents and 150 to 500 liquidity taker (i.e., market order placing) agents.

\citet{Challet:2003} studied how allowing the parameters of a simple ABM of an LOB to vary in time affected the traded price series, concluding that it resulted in the emergence of a heavy-tailed distribution of mid-price changes, autocorrelated mid-price returns, and volatility clustering.  They noted that many LOB models assume that parameter values remain constant in time, and they conjectured that several stylized facts (see Section \ref{stylizedfacts}) might be caused by real traders changing their actions over time.

\citet{Lillo:2007limit} showed how an ABM could explain the empirically observed power-law distribution of relative prices of incoming orders (see Section \ref{relpricearrivals}).  In particular, he solved a utility maximization problem to show that if mid-price movements were assumed to follow a Brownian motion, then each perfectly rational agent should choose the relative price of their submitted orders to be\begin{equation}\label{lilloeq}
{\delta^x}^{*} = \sqrt{2}g^{-1}(\alpha)V T^{\frac{1}{2}},
\end{equation}where $g(\alpha)$ describes the agent's risk aversion, $T$ is the agent's maximum time horizon (i.e., the maximum length of time that the agent is willing to wait before performing the trade), and $V$ is the market volatility.  He then studied how empirically observed homogeneity in $g$ and $T$ and fluctuations in $V$ affectted the price choices of interacting agents with different risk aversions $g$ and different maximum time horizons $T$.  He concluded that heterogeneity in $T$ was the most likely source of the power-law tails in the distribution of ${\delta^x}^{*}$, and that the homogeneity in $g$ and fluctuations in $V$ that have been observed empirically in a wide range of markets were unlikely to lead to a power-law tail in the distribution of ${\delta^x}^{*}$.

\section{Key Unresolved Problems}\label{openproblems}

In this section, we discuss key unresolved problems currently facing researchers of LOBs.

\begin{figure}
\centering
\includegraphics[width=9 cm]{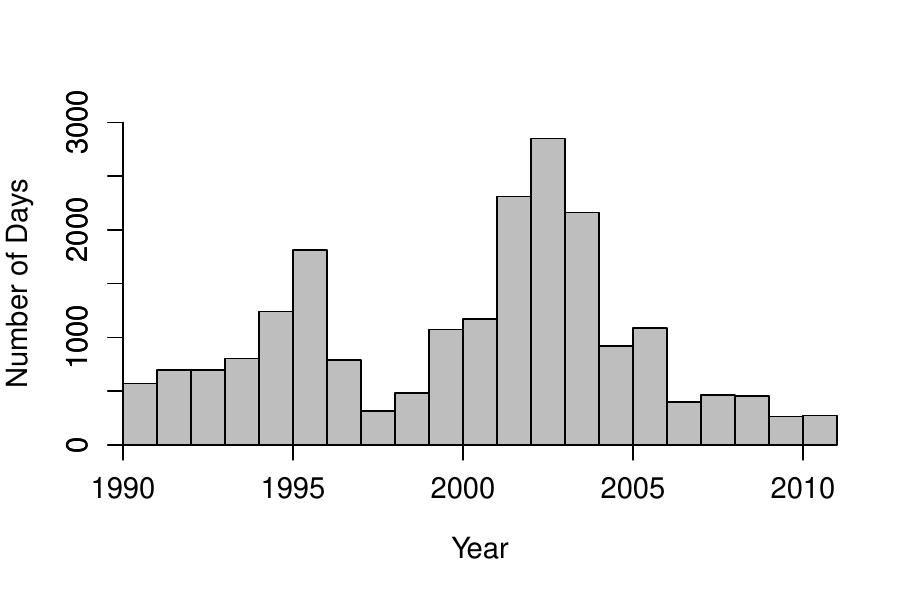}
\caption{Approximate total number of days' data per year that has been examined by empirical studies discussed in this article.}
\label{studiesplot}
\end{figure}
\begin{itemize}
\item \textbf{Understanding statistical regularities:} As discussed in Section \ref{empiricals}, several empirical regularities appear in data from a wide range of different markets.  Some such statistical regularities describe features of order flow or LOB state; others describe stylized facts relevant to price formation and market stability.  Many authors (see, e.g., \citet{Gu:2009, Lillo:2007limit, Stanley:2008statistical, Toth:2011anomalous}) agree that one of the main challenges facing researchers of LOBs is to gain a better understanding the origins of these statistical regularities.  LOB models can help to achieve this, and some progress has been made.  However, no single model has yet been capable of simultaneously reproducing all of the statistical regularities, and there is no clear picture about how the stylized facts emerge as a consequence of the actions of many heterogeneous traders.
\item \textbf{Understanding recent data:} A great deal of effort has been invested in empirical study of LOB data.  Figure \ref{studiesplot} shows the approximate number of days' data per year that studies discussed in this article have examined.  Although the breadth of such empirical work is substantial, the overwhelming picture painted by Figure \ref{studiesplot} is that the data studied is old.  It is also often of poor quality, so extensive auxiliary assumptions are required before any statistical analysis can even begin.  Strong assertions have been made by empirical studies based on single stocks over very short time periods.  Many LOB models are built upon statistical regularities observed in old data, but traders' strategies and the rules governing trade change over time, so empirical observations from more than a decade ago may not accurately describe current LOB activity.  However, recent advances in computational and storage capabilities have made it feasible to record data detailing all order flows at all prices, and tools have been developed to assist researchers with reconstructing the full LOB in certain markets \citep{Huang:2011LOBSTER}.  By studying recent, high-quality data, researchers will be able to assess whether the existing foundations for LOB modelling accurately reflect today's markets.
\item \textbf{Nonstationary behaviour:} Although precisely what is meant by ``equilibrium'' depends upon context, almost all LOB models to date have focused on some form of equilibrium, such as a Markov-perfect equilibrium in sequential-game models or a state-space equilibrium in reaction-diffusion models.  However, empirical evidence strongly suggests that LOBs are subject to frequent shocks in order flow that cause them to display nonstationary behaviour, so they may never settle into equilibrium \citep{Buchanan:2009waiting}.  Preliminary work on nonequilibrium models has hinted at promising results, such as quantitative replication of some of the stylized facts \citep{Challet:2003}, but there is very little progress in this direction.
\item \textbf{Volatility:} Price changes and volatility are among the most hotly debated topics in the literature \citep{Almgren:2001, Bouchaud:2009, Hasbrouck:1991, Potters:2003, Toth:2011anomalous}.  How can estimates of volatility be designed to incorporate information about the entire state of $\mathcal{L}(t)$?  What causes volatility to vary over time?  Why should periods of high activity cluster together?  Why should price fluctuations be so frequent and so large on intra-day timescales, given that external news events occur so rarely \citep{Maslov:2000}?  It is not even agreed whether the number of market orders \citep{Jones:1994}, the size of market orders \citep{Gallant:1992}, or the fluctuation of liquidity \citep{Bouchaud:2009} plays the dominant role in determining volatility.  It seems likely that the answers to such questions will not be found in isolation, but rather that there is an intricate interplay between the many pieces of the volatility puzzle.  Recent work has attempted to tie together some of these ideas.  For example, \citet{Bouchaud:2009} and \citet{Wyart:2008} conjectured that volatility might be understood better by considering the need for traders to minimize their market impact.
\item \textbf{Algorithmic trading:} Electronic trading algorithms are able to process vast quantities of LOB data to interpret market conditions and submit or cancel orders in a small fraction of the time that it would take a human to perform the same task.  The use of electronic trading algorithms has increased rapidly in recent years, but empirical research in this area is extremely difficult due to a lack of data in which algorithmic trades are identified \citep{Chaboud:2011rise}.  To date, the published literature on algorithmic trading consists of only a handful of empirical studies and models, yet there is fierce debate about whether such algorithms are beneficial or detrimental to markets.  Different studies have drawn contradictory conclusions.  \citet{Chaboud:2011rise} and \citet{Hendershott:2011} reported that electronic trading algorithms narrow spreads, reduce adverse selection, speed up price discovery, increase liquidity, and improve the informativeness of $b(t)$ and $a(t)$.  However, \citet{Biais:2011equilibrium} and \citet{Kirilenko:2011flash} reported that electronic trading algorithms increase adverse selection, create an unfair advantage for wealthier traders, decrease liquidity, and exacerbate volatility during stressed market scenarios.  From a regulatory standpoint, it is crucial to understand how electronic trading algorithms affect market stability, yet almost nothing concrete is currently known.
\item \textbf{Liquidity fragmentation:} In recent years, it has become increasingly common for assets to be traded on several different electronic trading platforms simultaneously \citep{Bennett:2006market}.  The resulting competition between exchanges has stimulated technological innovation and driven down the fees incurred by traders, but it has also caused a fragmentation of liquidity because limit orders for a given asset are now spread between several different LOBs.  This poses a problem for empirical research, as the study of any individual LOB in isolation no longer provides a snapshot of the whole market for an asset.  Furthermore, differences between different trading platforms' matching rules and transaction costs complicate comparisons between different LOBs for the same asset.  \citet{Cont:2011} reported similarities between different LOBs that traded the same asset simultaneously, but there is no reason that this must hold in general.  The development of robust methods for assimilating data across multiple platforms will be of paramount importance to understand the implications of liquidity fragmentation on market stability and price formation.
\end{itemize}

\section{Conclusion}\label{concl}

The literature on LOBs has grown rapidly, and both empirical and theoretical work has deepened understanding of the LOB trading process.  LOBs are a rich and exciting testing ground for theories, and have provided new insight into longstanding economic questions regarding market efficiency, price formation, and the rationality of traders.  However, despite the progress made on specific aspects of limit order trading, it remains unclear how the various pieces of the puzzle fit together.  For example, models that capture the dynamics of event-by-event price changes poorly reproduce price dynamics on longer timescales.  Similarly, models that explain price dynamics on inter-day timescales offer little understanding of how they emerge as the aggregate effect of individual trades.

There are substantial challenges associated with studying historical LOB data, and several empirical studies contain systematic errors in their calculations.  Moreover, performing quantitative comparisons between different empirical studies is very difficult for two reasons.  First, it is unclear whether differences in the findings of such studies are caused by differences in different markets, or are simply a result of differences in methodology.  Sampling frequency, choice of asset class, LOB resolution parameters, specific trade-matching nuances, and many other factors all influence empirical findings, but so too do the choice of statistical estimators and the details of their implementation.  This makes it difficult to assess the influence of specific LOB factors on trade.  Second, LOB platforms, LOB rules, and trading strategies have all changed over time, so the date range over which data was collected may itself play a role in the values of the statistics reported.  This issue is particularly important given the recent surge in popularity of electronic trading algorithms.  Studies of recent, high-quality LOB data that are conducted with stringent awareness of potential statistical pitfalls are needed to understand better the LOBs of today.

It is clear from empirical studies how poorly the data supports the very strong assumptions made by many LOB models.  Although every model must make assumptions to facilitate computation, many LOB models depend on elaborate and inaccurate assumptions that make it almost impossible to relate their output to real markets.  ABMs appear to offer some compromise between the extremes of zero-intelligence and perfect-rationality models; they also provide an explicit way to remove the inherent homogeneity associated with many existing approaches \citep{Lux:2009economics, Toke:2011, Zhao:2010}.  Furthermore, the level of game-theoretic considerations involved in agents' decision-making can be controlled by specifying how strongly agents react to each other and forecast each other's actions.  Therefore, ABMs have the potential to provide a rich toolbox for investigating LOBs without requiring extreme modelling assumptions.  However, it remains unclear whether the ABMs studied to date offer a deeper understanding of market dynamics or merely amount to curve-fitting exercises in which parameters are varied until some form of nontrivial behaviour emerges.  Recently, statistical techniques from econometrics have enabled consistent estimation of ABMs' parameters from market data \citep{Chen:2012agent}.  It will be interesting to see whether the use of such techniques in an LOB context paves the way for new, quantitative explanations of LOB phenomena.

Price impact and market impact also continue to be active areas of research.  A deeper understanding of these notions is very desirable, as they form a conceptual bridge between the microeconomic mechanics of order matchings and the macroeconomic concepts of price formation.  Considerations about price impact and market impact could also help to explain the actions of traders in certain situations.  However, despite the striking regularities that have been observed in empirical studies, little is understood about why price impact functions have the forms that they do and almost nothing is understood about market impact.  However, this is a difficult task because the state space of $\mathcal{L}(t)$ is so large.

LOBs have revolutionized trading by providing traders the freedom to evaluate their own need for immediate liquidity.  Their study has hitherto been hampered by their inherent complexity, with all the associated technical difficulties, and above all by wholly inadequate empirical data.  However, our growth in understanding allied to massive improvements in data and in computational power suggest that answers to many important open questions will not be long in coming.

\section*{Acknowledgements}We thank Bruno Biais, Jean-Philippe Bouchaud, Doyne Farmer, Gabriele La Spada, Sergei Maslov, Stephen Roberts, Torsten Sch\"{o}neborn, Cosma Shalizi, Neil Shephard, Eric Smith, Jonathan Tse, Thaleia Zariphopoulou, and Wei-Xing Zhou for useful discussions.  We also thank two anonymous reviewers for many helpful comments and suggestions.  Martin Gould and Sam Howison thank the Oxford-Man Institute of Quantitative Finance, and Martin Gould thanks EPSRC (Industrial CASE Award 08001834) and HSBC Bank, for financial support.

\bibliographystyle{rQUF}
\bibliography{limitorderbooksbib}

\clearpage

\appendix
\section{Table of Empirical Studies}\label{empiricaltable}
\begin{longtable}{|p{0.8in}|p{1.4in}|p{1.0in}|p{1.6in}|p{1.9in}|}
\hline
\textbf{Reference}&\textbf{Assets Studied}&\textbf{Date Range}&\textbf{Data Type}&\textbf{Main Points Studied}\\
\hline
\citet{AS:2011ultra}&The 30 Dow Jones Industrial Average stocks&19th -- 23rd and 26th -- 30th Apr 2004&$b(t)$, $a(t)$, $n^b(b(t),t)$, $n^a(a(t),t)$, and all market orders&Volatility and long range dependence in order flows \\
\hline
\citet{Anand:2005}&144 stocks traded on the NYSE&Nov 1990 -- Jan 1991&All order flows at all prices&Decision between using limit orders or market orders for informed traders \\
\hline
\citet{Bandi:2006}&All stocks in the S\&P 100 index&Feb 2002&$b(t)$, $a(t)$, $n^b(b(t),t)$, $n^a(a(t),t)$, and all market orders&Volatility \\
\hline
\citet{Bennett:2006market}&39 stocks that voluntarily switched their listing from NASDAQ to the NYSE&Jan 2002 -- Mar 2003&Total size of arriving market orders, daily returns, $s(t)$, and several summary statistics describing individual assets&How liquidity fragmentation affects markets \\
\hline
\citet{Biais:1995}&The CAC 40, traded on the Paris Bourse&6 trading days in Jun/Jul 1991 and 19 trading days in Oct/Nov 1991&First 5 levels of bid-side relative depth profile and ask-side relative depth profile (updated every time the depth available at one of the displayed levels changed)&Returns, percentage of market orders that matched to hidden liquidity, mean total depth available, and $s(t)$ (both unconditionally and dependent on time of day), order flow (both unconditionally and dependent on recent order flow and time of day), and state of $\mathcal{L}(t)$ \\
\hline
\citet{Biais:1999}&The CAC 40, traded on the Paris Bourse&19 trading days in Oct/Nov 1991, 26 trading days in 1993, and 234 trading days in 1995&$b(t)$ and $a(t)$ (sampled once per minute)&Whether the evolution of the price process indicates that traders learn during the daily opening auction\\
\hline
\citet{Boehmer:2005}&400 stocks traded on the NYSE&7th -- 18th Jan, 4th -- 15th Feb, 4th -- 15th Mar, 1st -- 12th Apr, 6th -- 17th May (all in 2002)&All order flows at all prices in the electronic LOB and information about the handling of both electronic and manual (broker-handled) orders&How the introduction of an electronic LOB on the NYSE affected traders' behaviour \\
\hline
\citet{Bortoli:2006}&The 4 most actively traded futures contracts on the Sydney Futures Exchange&15th Sep 2000 -- 19th Jun 2001&Every matching, change in $b(t)$ or $a(t)$, and change in depth available at the best prices (respectively, best three prices) prior to (respectively, after) the change in disseminated market information (timestamped to the nearest second)&Whether order flow and $\mathcal{L}(t)$ changed when the Sydney Futures Exchange increased the real-time information disseminated to traders \\
\hline
\citet{Bouchaud:2002}&France Telecom, Vivendi, and Total stocks, traded on the Paris Bourse&Feb 2001&All order arrivals at all prices along with their time of arrival and a list of all orders that were cancelled (but not the time at which they were cancelled)&Mean depths available, distribution of relative prices, $\omega_x$, $n^b(b(t),t)$, and $n^a(a(t),t)$ \\
\hline
\citet{Bouchaud:2004}&France Telecom Stock, traded on the Paris Bourse (with similar results reported for other unnamed liquid French and British stocks)&Jan 2001 -- Dec 2002&$b(t)$ and $a(t)$ (recorded once every time either of them changed) and all market orders (timestamped to the nearest second)&How order flow affects prices \\
\hline
\end{longtable}
\clearpage
\begin{longtable}{|p{0.8in}|p{1.4in}|p{1.0in}|p{1.6in}|p{1.9in}|}
\hline
\citet{Cao:2008}&100 largest stocks traded on the Australian Stock Exchange&Mar 2000&All order arrivals and cancellations at all prices (timestamped to the nearest $0.01$ seconds)&How the state of $\mathcal{L}(t)$ affects order flow \\
\hline
\citet{Chaboud:2011rise}&EUR/USD, USD/JPY, and EUR/JPY currency pairs on EBS&2003 -- 2007&$b(t)$ and $a(t)$ (sampled once per second) and total size of arriving market orders (sampled once per minute)&How electronic trading algorithms affect markets \\
\hline
\citet{Chakraborti:2011a}&Four stocks traded on the Paris Bourse&All trading days, 1st Oct 2007 -- 30th May 2008&All market orders and the five highest-priority active orders on each side of the LOB&Whether the traditional stylized facts are present in the data \\
\hline
\citet{Challet:2001}&Four stocks traded on the Island ECN (on NASDAQ)&Not specified&15 highest-priority active orders on each side of the LOB (updated every time the list changed)&Order flow rates, autocorrelation of order flow rates, diffusion of active orders (i.e., cancellation of an active order immediately followed by resubmission at a neighbouring price), instantaneous price impact, distribution of $\omega_x$, lifetime of limit orders, and $\delta^x$ for incoming orders \\
\hline
\citet{Cont:2010}&Sky Perfect Communications stock, traded on the Tokyo Stock Exchange&Not specified&$N^b(p,t)$ and $N^a(p,t)$ for the five smallest relative prices with nonzero depth available (updated whenever either changed) and all market orders&Arrival rates of market orders and arrival and cancellation rates of limit orders \\
\hline
\citet{Cont:2011}& 50 stocks from the S\&P 500, traded on the NYSE&All 21 trading days in Apr 2010&$n^b(b(t),t)$ and $n^a(a(t),t)$ (updated whenever either changed and timestamped to the nearest second) and all market orders&Relationship between order flow imbalance and price impact \\
\hline
\citet{Dufour:2000}&18 of the most frequently traded stocks on the NYSE&62 trading days, 1st Nov 1990 -- 31st Jan 1991&$b(t)$ and $a(t)$(updated whenever either changed) and all market orders&Relationship between market order inter-arrival times and price impact \\
\hline
\citet{Eisler:2012}&14 stocks traded on NASDAQ&3rd Mar 2008 -- 19th May 2008&$b(t)$, $a(t)$, $n^b(b(t),t)$, and $n^a(a(t),t)$(updated whenever any of them changed)&Price impact of market order submissions, and limit order submissions and cancellations \\
\hline
\citet{Ellul:2003}&The 50 most actively traded stocks and 98 other stocks on the NYSE&30th Apr 2001 -- 5th May 2001&All market order submissions and all limit order submissions and cancellations (timestamped to the nearest second)&Which factors traders assess when choosing the price of an order \\
\hline
\citet{Engle:2004}&100 stocks traded on the NYSE&18 months of data, no specified date range&$b(t)$ and $a(t)$ (updated whenever either changed) and all market orders&$s(t)$ and how price impact varies according to how frequently trades occur for a specific stock\\
\hline
\citet{Farmer:2003}&3 stocks traded on the LSE and 3 stocks traded on the NYSE&May 2000 -- Dec 2002 for the LSE stocks and 1995 -- 1996 for the NYSE stocks&All order flows for the LSE; $b(t)$ and $a(t)$ (updated whenever either changed) and all market orders for the NYSE&Price impact of individual market orders and distribution of $\omega_x$ for market orders \\
\hline
\citet{Farmer:2005}&11 stocks traded on the LSE&1st Aug 1998 -- 30th Apr 2000&All market order submissions and all limit order submissions and cancellations&Goodness-of-fit of the predictions regarding mean spread and price diffusion of the \citet{Smith:2003} model and mean instantaneous mid-price logarithmic return impact as a function of market order size \\
\hline
\citet{Field:2008}&Short Sterling, Euribor, EUR/USD, and 2-Year US Treasury Note futures&23rd Nov -- 11th Dec 2006 and 16th -- 20th Apr 2007&$b(t)$, $a(t)$, $n^b(b(t),t)$, and $n^a(a(t),t)$ (updated whenever any of them changed)&Order flow rates and $n^b(b(t),t)$ and $n^a(a(t),t)$ in markets in which $s(t)=\pi$ \\
\hline
\end{longtable}
\clearpage
\begin{longtable}{|p{0.8in}|p{1.4in}|p{1.0in}|p{1.6in}|p{1.9in}|}
\hline
\citet{Gode:1993}&Laboratory experiment with human beings and computerized zero-intelligence traders&N/A&All order flows at all prices&Relative applicability of perfect-rationality and zero-intelligence assumptions, and emergence of seemingly rational behaviour when aggregating across irrational individuals \\
\hline
\citet{Gopikrishnan:2000}&1000 largest stocks traded in the US&1994 -- 1995&$a(t)$, $b(t)$, and all market orders&Price impact as a function of trade imbalance count and trade imbalance size, and distribution and autocorrelation of trade imbalance count and trade imbalance size \\
\hline
\citet{Gu:2008empirical}&23 stocks traded on the Shenzhen Stock Exchange&All of 2003&All order flows at all prices&Distribution of mid-price returns on various $\tau$ second timescales and various event-by-event timescales \\
\hline
\citet{Gu:2008empiricalregularities}&23 stocks traded on the Shenzhen Stock Exchange&All of 2003&All order flows at all prices&Distribution of relative prices of incoming orders and whether this is conditional on $s(t)$ or volatility \\
\hline
\citet{Gu:2008empiricalshape}&23 stocks traded on the Shenzhen Stock Exchange&All of 2003&All order flows at all prices&$\overline{N}^b(p)$, $\overline{N}^a(p)$, and changes in relative depth profiles through time\\
\hline
\citet{Gu:2009}&23 stocks traded on the Shenzhen Stock Exchange&All of 2003&All order flows at all prices&Autocorrelation of $\delta^x$ for incoming orders \\
\hline
\citet{Hall:2006}&The 5 most liquid stocks traded on the Australian Stock Exchange&Jul -- Aug 2002&All order flows at all prices&Whether the distribution of $\delta^x$ for incoming orders is conditional on $\mathcal{L}(t)$, volatility, and recent order flows \\
\hline
\citet{Harris:1996}&144 stocks traded on the NYSE&Nov 1990 -- Jan 1991&All order flows at all prices&Analysis of performance measures aiding decision-making between limit orders versus market orders \\
\hline
\citet{Hasbrouck:2002}&The 300 largest equities on NASDAQ, traded on Island ECN&1st Oct -- 31st Dec 1999&All order flows at all prices&How volatility is related to order flow and $\mathcal{L}(t)$, and how order fill probabilities and mean time to execution vary with volatility \\
\hline
\citet{Hautsch:2011market}&The 30 most frequently traded stocks on Euronext Amsterdam &All trading days between 1st Aug and 30th Sep, 2008&First 2 levels of bid-side relative depth profile and ask-side relative depth profile (updated whenever either changed) and a record of all trades (timestamped to the nearest millisecond)&Market impact of incoming limit orders \\
\hline
\citet{Hendershott:2005}&3 exchange-traded funds on Island ECN&16th Aug -- 31st Oct 2002&For activity on Island: for first part of data, $b(t)$, $a(t)$, $n^b(b(t),t)$, and $n^a(a(t),t)$ (updated whenever any of them changed), and all market orders; for second part of data, only market orders; for activity not on Island, $b(t)$, $a(t)$, $n^b(b(t),t)$, and $n^a(a(t),t)$ (updated whenever any of them changed), and all market orders for entire data period&How showing traders $\mathcal{L}(t)$ affects price series \\
\hline
\citet{Hendershott:2011}&943 stocks traded on the NYSE&Feb 2001 -- Dec 2005&$b(t)$, $a(t)$, $n^b(b(t),t)$, and $n^a(a(t),t)$ (updated whenever any of them changed)&How algorithmic trading affects $\mathcal{L}(t)$ \\
\hline
\citet{Hollifield:2004}&The Ericsson stock, traded on the Stockholm Stock Exchange&59 trading days, 3rd Dec 1991 -- 2nd Mar 1992&All order flows at all prices&Whether traders' actions can be explained by a cut-off strategy based on their private valuation of the traded asset \\
\hline
\end{longtable}
\clearpage
\begin{longtable}{|p{0.8in}|p{1.4in}|p{1.0in}|p{1.6in}|p{1.9in}|}
\hline
\citet{Hollifield:2006}&3 stocks traded on the Vancouver Stock Exchange&May 1990 -- Nov 1993&All order flows at all prices&Distribution of traders' personal valuations (inferred from their actions) \\
\hline
\citet{Kempf:1999}&DAX futures contracts, traded on the German Futures and Options Exchange&17th Sep 1993 -- 15th Sep 1994&$b(t)$, $a(t)$, and all market orders&Permanent price impact, as a function of several measures of trade imbalance, over $1$-minute time horizons \\
\hline
\citet{Kirilenko:2011flash}&E-mini S\&P 500 index futures contracts, traded on the Chicago Mercantile Exchange&6th May 2010&All order flows at all prices, including details of which traders submitted which orders&Possible causes of the ``Flash Crash'' \\
\hline
\citet{Lillo:2004long}&20 stocks traded on the LSE&1999 -- 2002&All order flows at all prices&Autocorrelations of $\omega_x$ series, $m(t)$, $n^b(b(t),t)$, $n^a(a(t),t)$, and order type (buy or sell) for arriving LOs, arriving MOs, and cancelled LOs \\
\hline
\citet{Lillo:2005}&20 stocks traded on the LSE&May 2000 -- Dec 2002&All LOB order flows and all off-book trades for the same stocks&Effects of order splitting and hidden liquidity on observed order flows \\
\hline
\citet{Lillo:2007limit}&Astrazeneca Stock, traded on the LSE&May 2000 -- Dec 2002&Order arrivals, partitioned by who submitted them&Distribution of $\delta^x$ for incoming limit orders from specified traders \\
\hline
\citet{Lo:2010}&DEM/USD and USD/CAD currency pairs, traded on Reuters&5th Oct -- 10th Oct 1997 for DEM/USD; 1st May -- 30th June 2005 for USD/CAD&All order flows at all prices&How traders choose $\omega_x$ and $\delta^x$ for their orders \\
\hline
\citet{Madhavan:2005}&109 stocks traded via LOBs and 240 stocks traded by floor traders on the Toronto Stock Exchange&Mar and May, 1990&For March: $b(t)$, $a(t)$, $n^a(a(t),t)$, and $n^b(b(t),t)$; for May: $b(t)$, $a(t)$, first 5 levels of bid-side relative depth profile and ask-side relative depth profile; for both months: all market orders all floor-trader trades&How real-time disclosure of more information about the depth profile affects traders' behaviour \\
\hline
\citet{Maskawa:2007correlation}&13 stocks traded on the LSE&Jul -- Dec 2004&All order flows at all prices&Distribution of $\delta^x$ for incoming limit orders, and whether this distribution is affected by the state of $\mathcal{L}(t)$\\
\hline
\citet{Maslov:2001}&Cisco Systems, Broadcom Corporation, and JDS Uniphase Corporation stocks (traded on NASDAQ)&30th Jun 2000 for Cisco Systems; 3rd Jul for Broadcom Corporation; and 5th, 6th, and 11th Jul for JDS Uniphase Corporation&$b(t)$, $a(t)$, first 4 levels of bid-side relative depth profile and ask-side relative depth profile, and all market orders&Distribution of $\omega_x$, $n^b(b(t),t)$, $n^a(a(t),t)$, depth profiles, and instantaneous price impact\\
\hline
\citet{Mike:2008}&25 stocks traded on the LSE&May 2000 -- Dec 2002&All order flows at all prices&$\delta^x$ for incoming orders, autocorrelation of order type in order flows, and order cancellations \\
\hline
\citet{Mizrach:2008}&The 4 largest stocks on NASDAQ; 95 of the NASDAQ 100 stocks; and 87 other smaller NASDAQ stocks&Dec 2002&All order flows at all prices&How $\mathcal{L}(t)$ affects the next change in $b(t)$ or $a(t)$ \\
\hline
\citet{Mu:2009preferred}&22 stocks traded on the Shenzhen Stock Exchange&All of 2003&All order flows at all prices&Distribution of $\omega_x$ for market orders \\
\hline
\end{longtable}
\clearpage
\begin{longtable}{|p{0.8in}|p{1.4in}|p{1.0in}|p{1.6in}|p{1.9in}|}
\hline
\citet{Mu:2010tests}&978 stocks traded on the Shenzhen Stock Exchange&Jan 2004 -- Jun 2006&$b(t)$ and $a(t)$ (updated once every 6 -- 8 seconds)&Distribution of mid-price logarithmic returns for stocks in emerging markets, and how this varies with time window and market capitalization of the studied asset \\
\hline
\citet{Plerou:2002}&The 116 most frequently traded US stocks&1994 -- 1995&$b(t)$, $a(t)$, $n^b(b(t),t)$, $n^a(a(t),t)$, and all market orders&Price impact as a function of trade imbalance count and trade imbalance size over a variety of time horizons \\
\hline
\citet{Plerou:2008stock}&1000 major US stocks; 85 of the FTSE 100 stocks, traded on the LSE; 13 of the CAC 40 stocks, traded on the Paris Bourse; 422 stocks from the Center for Research in Security Prices (CRSP)&1994 -- 1995 for US stocks; 2001 -- 2002 for LSE stocks; 3rd Jan 1995 -- 22nd Oct 1999 for Paris Bourse stocks; Jan 1962 -- Dec 1996 for CRSP database stocks&All market orders&Distribution of mid-price returns and number of arriving market orders, and whether they vary according to market capitalization or industry sector, on various $\tau$-second timescales \\
\hline
\citet{Potters:2003}&Exchange-traded funds that track NASDAQ and the S\&P 500, and the Microsoft stock&1st Jun -- 15th Jul, 2002&All order flows at all prices&Distribution of $\delta^x$ for arriving orders, relative depth profiles, arrival and cancellation rates, and instantaneous price impact \\
\hline
\citet{Ranaldo:2004order}&15 stocks traded on the Swiss Stock Exchange&Mar and Apr 1997&$b(t)$, $a(t)$, $n^b(b(t),t)$, $n^a(a(t),t)$, and all market orders&How volatility, recent order flow, and the state of $\mathcal{L}(t)$ affect order flow, intra-day patterns in $s(t)$ and volatility, symmetry between the buy side and sell side \\
\hline
\citet{Sandas:2001}&10 stocks traded on the Stockholm Stock Exchange&3rd Dec 1991 -- 2nd Mar 1992&All order flows at all prices&Whether the depth profile supports hypotheses about how traders make decisions related to order submissions and cancellations \\
\hline
\citet{Toke:2011}&3 stocks from the CAC 40, 3 month Euribor futures, and FTSE 100 futures&10th Sep 2009 -- 30th Sep 2009&First 5 levels of bid-side relative depth profile and ask-side relative depth profile (updated whenever any of them changed and timestamped to the nearest millisecond)&Whether Hawkes processes provide a better explanation of order flows than do Poisson processes \\
\hline
\citet{Toth:2011anomalous}&500,000 trades on a variety of futures contracts&Jun 2007 -- Dec 2010&Changes in $b(t)$ and $a(t)$&Price impact \\
\hline
\citet{Wyart:2008}&The 68 most liquid stocks on the Paris Bourse, small tick index futures contracts, and the 155 most actively traded stocks on the NYSE&2002 for the Paris Bourse and 2005 for the small tick futures and NYSE stocks&$b(t)$, $a(t)$, $n^b(b(t),t)$ and $n^a(a(t),t)$, and all market orders&Price impact and how the profit of a market maker trading in an LOB depends on $s(t)$ \\
\hline
\citet{Zhao:2010}&Crude oil futures contracts, traded on the International Petroleum Exchange&17th Oct 2005&First 5 levels of bid-side relative depth profile and ask-side relative depth profile (updated whenever either changed) and all market orders (timestamped to the nearest second)&Order flow rates \\
\hline
\end{longtable}
\clearpage
\begin{longtable}{|p{0.8in}|p{1.4in}|p{1.0in}|p{1.6in}|p{1.9in}|}
\hline
\citet{Zhou:1996}&DEM/USD, USD/JPY and DEM/JPY currency pairs, traded on Reuters&1st Oct 1992 -- 30th Sep 1993&$b(t)$&Volatility \\
\hline
\citet{Zhou:2012universal}&23 stocks traded on the Shenzhen Stock Exchange (although 1 is later removed, as its price was reported to be manipulated in the data)&All of 2003&All order flows at all prices&Instantaneous price impact of individual orders \\
\hline
\end{longtable}

\end{document}